\documentclass{jfm}
\usepackage[version=3]{mhchem}
\usepackage{subfigure}
\usepackage{color}
 \usepackage{dblfloatfix}
 \usepackage{graphicx}

 \usepackage{citesort}
\usepackage{ amssymb }
\usepackage{setspace}
\usepackage{siunitx}
\usepackage{hyperref}
\DeclareMathAlphabet{\mathpzc}{OT1}{pzc}{m}{it}

\usepackage{siunitx}
\title{Asymptotic theory for a Leidenfrost drop on a liquid pool}
\author{Michiel A.J. van Limbeek\aff{1}
  \corresp{\email{m.a.j.vanlimbeek@utwente.nl}},
  Benjamin Sobac\aff{2},  
  Alexey Rednikov\aff{2},
  Pierre Colinet\aff{2}
 \and  Jacco H. Snoeijer\aff{1},
 }
 
\affiliation{\aff{1}University of Twente, Physics of Fluids group, P.O. Box 217 7500AE Enschede 
\aff{2}Universit\'e Libre de Bruxelles, TIPs-Fluid Physics, C.P. 165/67, Av. F.D. Roosevelt 50, 1050 Brussels, Belgium
}

\begin{document}
\maketitle
%\noindent\LARGE{\textbf{Unravelling the  dynamics of Leidenfrost drops on a liquid pool}}\\ \noindent\large{\textbf{M. A. J. van Limbeek,$^{\ddag}$\textit{$^{a}$} B. Sobac,\textit{$^{b}$} A. Rednikov,\textit{$^{b}$} D. Lohse\textit{$^{a}$}, P. Colinet,\textit{$^{b}$} and J. Snoeijer,\textit{$ ^{a}$}}}\\ \textit{$^{a}$ University of Twente, Physics of Fluids group, P.O. Box 217 7500AE Enschede } \\  \textit{$^{b}$ Universit\'e Libre de Bruxelles, TIPs-Fluid Physics, C.P. 165/67, av. F.D. Roosevelt 50, 1050 Brussels, Belgium}\\  $^{\ddag}$ corresponding author: m.a.j.vanlimbeek@utwente.nl\\
\begin{abstract}
Droplets can be levitated by their own vapour when placed onto a superheated plate (the Leidenfrost effect). It is less known that the Leidenfrost effect can likewise be observed over a liquid pool (superheated with respect to the drop), which is the study case here. Emphasis is placed on an asymptotic analysis in the limit of small evaporation numbers, which proves to be a realistic one indeed for not so small drops. The global shapes are found to resemble ``superhydrophobic drops" that follow from the equilibrium between capillarity and gravity. However, the morphology of the thin vapour layer between the drop and the pool is very different from that of classical Leidenfrost drops over a flat rigid substrate, and exhibits different scaling laws. We determine analytical expressions for the vapour thickness as a function of temperature and material properties, which are confirmed by numerical solutions. Surprisingly, we show that deformability of the pool suppresses the chimney instability of Leidenfrost drops. 
\end{abstract}

\section{Introduction}

\begin{figure}
\centering
\includegraphics[scale=0.5]{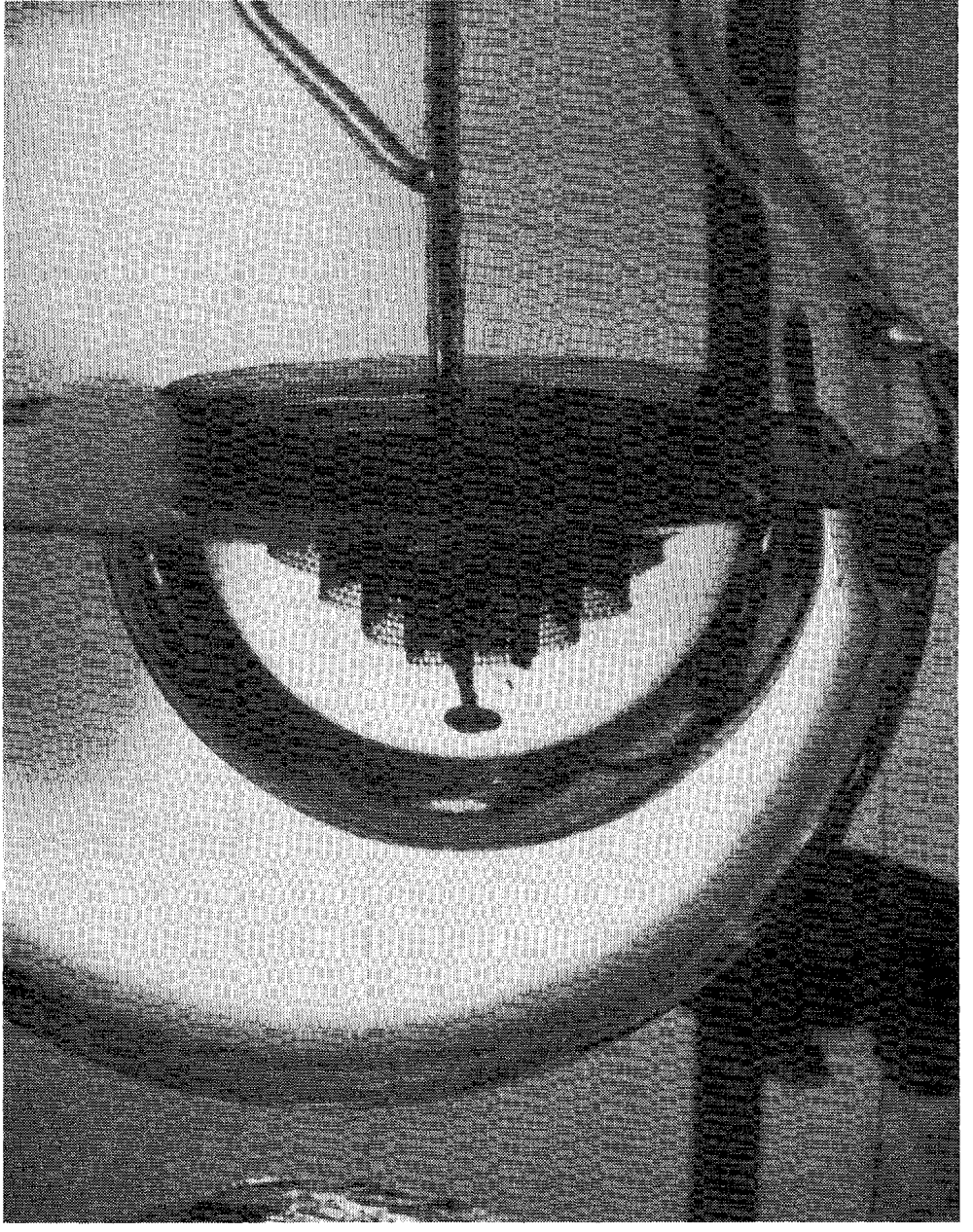}
\caption{A Leidenfrost drop of water floating on a water bath that is heated a few degrees above the saturation temperature. The drop, also referred to as ``boule" has a radius of \SI{7}{\centi\meter}, which is 25 times the capillary length.  (Reprinted with permission from   \citet{Hickman64} 'Floating drops and liquid boules'. Copyright 2016 American Chemical Society.)}
\label{fig:HickBoule}
\end{figure}

A drop can be prevented from merging with a liquid bath when the bath is heated above the saturation temperature. Recently, such Leidenfrost drops have regained attention  \citep{Maquet2016} after the first reports by \citet{Hickman64}, who more than half a century ago referred to these drops as ``boules". Figure~\ref{fig:HickBoule} gives an example of such a large water drop that is prevented from contacting a pool of water. The evaporation gives rise to a thin vapour layer between the drop and the pool, and the corresponding vapour flow induces a pressure that keeps the drop separated from the pool.

Naturally, one tries to compare these ``boules" to drops levitated above a heated plate. The latter have been studied in great detail \citep{Wachters1966a,Bernardin1999,Biance2003,Quere2013} since the first report by \citet{Leidenfrost1756} centuries ago. Since then various studies have focussed on features as shape oscillations \citep{Norman1952,Takaki1985,Strier2000,Snezhko2008,Brunet2011,Bouwhuis2013,Ma2017}, drop mobility on ratchets or gradients \citep{linke2006,wurger2011,lagubeau2011,Sobac2017}, dynamics during drop impacts \citep{Chandra1991,Tran2012,Shirota2016}, and the Leidenfrost temperature  \citep{Baumeister1973,vanLimbeek2016}. 

\begin{figure}
\includegraphics[scale=0.4]{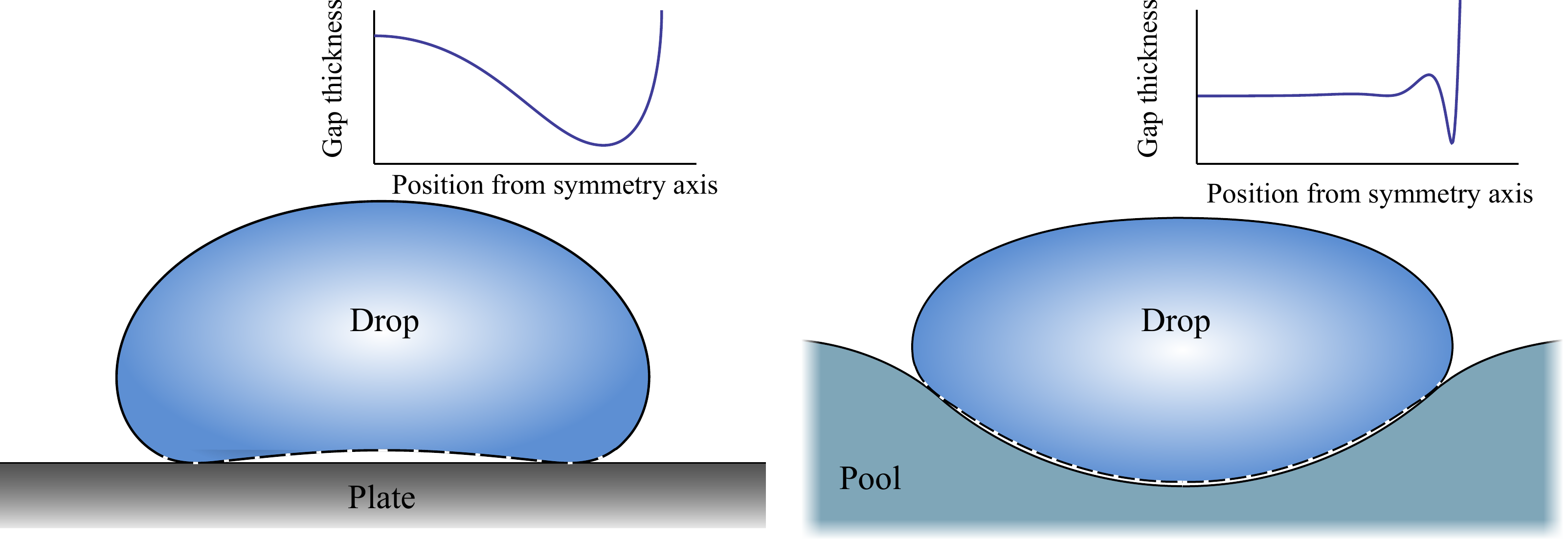}
\caption{Sketches of  a Leidenfrost drop levitated above a hot plate (left) and above a hot pool (right) based on numerical simulations. The resulting shapes are essentially ``superhydrophobic drops" (solid lines), underlied by a thin vapour layer (dashed lines). The insets provide a detailed zoom of the geometry of the vapour layer, revealing a striking difference between the two cases.}
\label{fig:Sketch}
\end{figure}

Of particular interest is the shape of such Leidenfrost drops. When viewed from the side, a Leidenfrost drop above a plate resembles a sessile drop that makes a contact angle of \SI{180}{\degree} with the substrate \citep{Biance2003,Snoeijer2009,Quere2013,Sobac2014}. The vapour layer prevents a direct contact so that the droplet is maintained in a perfectly non-wetting, ``superhydrophobic" state. Intriguingly, it is only fairly recently that the morphology of the thin vapour layer below the drop has been revealed. Experimentally, the shape was characterised by interferometry by \citet{Burton2012}, showing that the thickness of the vapour layer is not uniform. This is sketched in the left panel of \autoref{fig:Sketch}, where one observes a large vapour pocket near the center of the drop and a thin ``neck" near its edge. For large drop radii, the base of the drop can even penetrate up to the top enabling vapour to escape by a ``chimney instability" \citep{Biance2003,Snoeijer2009}. Even prior to experiments, the details of the layer below levitated drops were predicted from a hydrodynamic analysis  \citep{duchemin2005,Lister2008,Snoeijer2009}, with a more complete description of the evaporation developed by \citet{Sobac2014}. In the limit of small evaporation one indeed finds the neck to be asymptotically thinner than the vapour film at the center, and scaling laws characterizing such vapour pocket were established. One of the salient features is that the vapour pressure that carries the weight of the drop is nearly uniform below the drop: owing to the viscous resistance to vapour flow, the pressure falls abruptly across the thin neck to reach the atmospheric pressure.

Remarkably, Leidenfrost drops on a pool exhibit very different morphologies \citep{Maquet2016}. A typical numerical result is shown in the right panel of \autoref{fig:Sketch}. Globally the drop can still be considered in a ``superhydrophobic" state, but now on a deformable pool rather than on a flat substrate. The shape of the vapour layer, however, is very different from Leidenfrost drops on such a substrate: the vapour layer is nearly uniform and exhibits oscillations before passing a thin neck. Also,  the numerical analysis of \citep{Maquet2016} showed no indication of a chimney instability for Leidenfrost drops on a pool, even for drops considerably larger than the capillary length.

Note that the case by \citet{Maquet2016} is actually slightly different from  the boule case by \citet{Hickman64} in the following regard. In the former (Leidenfrost) case, the evaporative heat flux is limited by heat conduction across the vapour gap from the superheated pool surface (non-volatile) to the drop surface keeping at the saturation temperature, evaporation proceeding from the drop surface. In the latter (boule) case, the two liquids being the same, both surfaces of the vapour gap find themselves at the same, saturation temperature, while the evaporative heat flux is rather limited by heat transport from the superheated bulk of the pool with evaporation eventually taking place from the surface of the pool. 

In the present paper, a baseline consideration will explicitly be adapted to the former case \citep{Maquet2016}, where the boule case will be only mimicked by choosing equal densities and surface tensions of the two liquids. 

Our investigation of Leidenfrost drops on a pool will be based on a matched asymptotics analysis in the limit of small evaporation numbers. We compute the detailed structure of the vapour layer as in \autoref{fig:Sketch}, and establish the scaling laws for the thickness as a function of the material properties and the superheat. In \S\ref{sec:formulation} we formulate the problem and sketch the asymptotic structure,  which is worked out in detail in \S \ref{sec:analysis}. The boules of Hickman are discussed at the end of that section. Analytical results are obtained in the limit of large drops, explaining why, indeed, there is no chimney instability above a pool. The results are generalised in \S\ref{sec:extend} for the case of smaller drops and differing liquids, showing that the scaling laws are robust. The paper closes with a discussion in \S\ref{sec:Discussion}.

\section{Formulation}\label{sec:formulation}

In this section we first present a set of equations that describe a steady Leidenfrost drop levitated above a liquid pool. This part follows the ideas presented  in \citet{Maquet2016}, although the problem is formulated in terms more amenable to our present analysis. We then sketch how the equations are solved by means of matched asymptotic expansions. 
%We also provide a brief description of numerical implementation that is used to validate analytical results.

\subsection{Model}
\label{sec:model}
\begin{figure}
\centering
\includegraphics[scale=0.75]{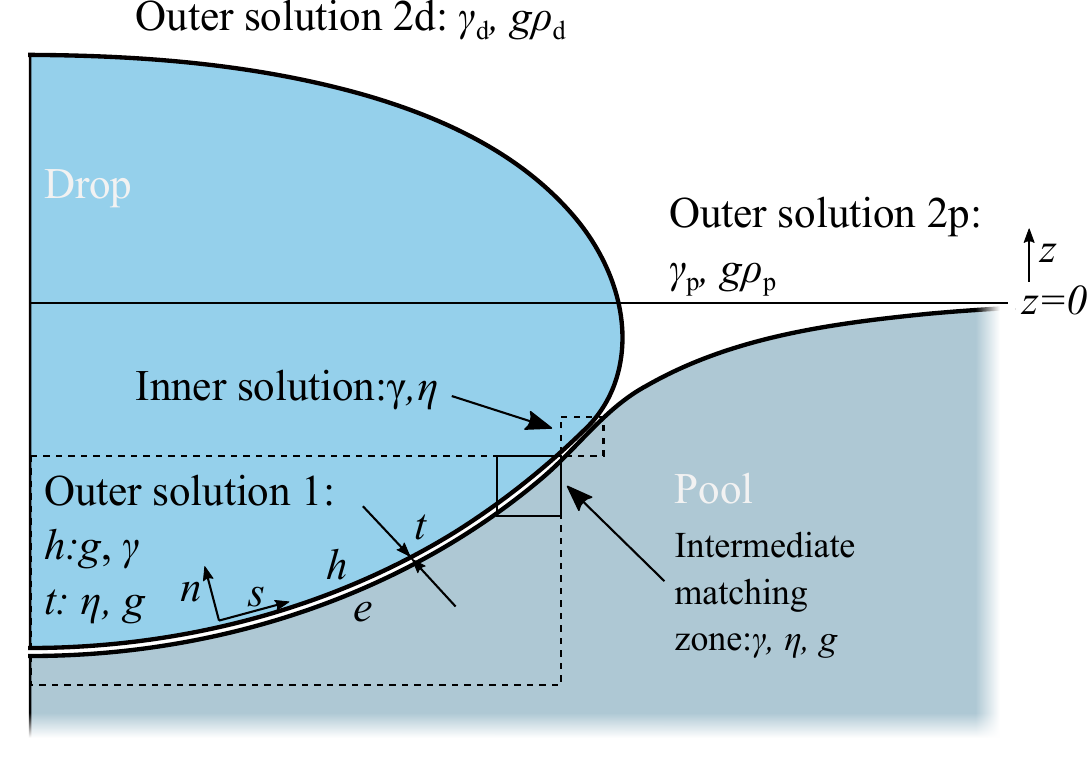}
\caption{Sketch of a drop on a pool. Different zones are identified which are used for the asymptotic analysis for ${\cal E} \ll 1$, ${\cal R} \gg 1$ for identical fluid properties ($\Gamma=1$, $\mathcal{P}=1$). The dominant force balances in the inner/outer regimes are indicated as capillary ($\gamma$), viscous ($\eta_\mathrm{v}$), and gravitational ($g$). Also defined are the vapour layer thickness $t$, which is the separation between the drop interface $h$ and the pool interface $e$, while $s$ and $n$ denote the orientation of the curvilinear coordinate system. Note that in the case of non-equal properties the thickness is also determined by capillary effects.}
\label{fig:sketchzones}
\end{figure}

We first need to establish a convenient representation of the drop-on-pool geometry shown in the right panel of figure~\ref{fig:Sketch}. The problem consists of two axisymmetric liquid domains, the drop and the pool, which we describe by the position of their respective liquid-vapor interfaces $z=h$ (for the drop) and $z=e$ (for the pool), where the $z$ axis points vertically upwards with $z=0$ corresponding to the unperturbed pool surface far away from the drop. These are defined in more detail in figure~\ref{fig:sketchzones}. While $h$ and $e$ in principle provide a full description of the geometry, we also introduce the thickness of the vapour layer $t$. This is convenient for describing the flow inside the thin vapour layer below the drop, where the two interfaces are essentially parallel. 

Following previous theoretical developments on levitated drops \citep{Lister2008,Snoeijer2009,Sobac2014,Maquet2016}, we consider the upper surfaces of the liquid drop and pool to be at hydrostatic equilibrium while the flow of the produced vapour is treated in the lubrication approximation. We first compute the vapour pressure $P_v$ by approaching it from the side of the pool,
\begin{equation}\label{eq:pvpool}
P_v=-\rho_\mathrm{p} g e+\gamma_\mathrm{p} \kappa_e.
\end{equation}
Here we introduced the pool density $\rho_\mathrm{p}$ and surface tension $\gamma_\mathrm{p}$, while the $\kappa_e$ is the curvature of the pool interface. The first term represents the hydrostatic pressure inside the pool, while the curvature term is the Laplace pressure jump due to surface tension. Note that the hydrostatic pressure inside the pool was taken $-\rho_\mathrm{p} g e$, i.e. the atmospheric pressure was set to zero. Similarly, we obtain an expression for the $P_v$ from the side of the drop
\begin{equation}\label{eq:pvdrop}
P_v=k-\rho_\mathrm{d} g h-\gamma_\mathrm{d} \kappa_h,
\end{equation}
where $\rho_\mathrm{d}$, $\gamma_\mathrm{d}$, $\kappa_h$ represent the droplet density, surface tension and curvature. Here k is a constant parametrising the size of the drop (see below). Equations (\ref{eq:pvpool},\ref{eq:pvdrop}) give two separate expressions for $P_v$, which in the lubrication approximation for thin layers must be identical. Therefore (\ref{eq:pvpool},\ref{eq:pvdrop}) can also be seen as a relation between $h$ and $e$.

We now turn to the flow inside the vapour layer. It will be shown that for sufficiently small evaporation rates the gap thickness $t$ is asymptotically small, justifying the use of the lubrication theory. We therefore consider the reduced Stokes equation for the parallel velocity $u$,
\begin{equation}
\partial_s P_\mathrm{v}=\eta_\mathrm{v} \partial_{nn} u,
\label{eq:Stokes}
\end{equation}
where $s$ is the curvilinear coordinate along the layer, while $n$ is the coordinate perpendicular to the vapour film (see figure~\ref{fig:sketchzones}). Owing to the small gas viscosity $\eta_\mathrm{v}$, we here assume that no flow is induced inside the drop and the pool. As a consequence, we can solve (\ref{eq:Stokes}) with no-slip boundary conditions at $n=0$ and $n=t$, yielding a parabolic profile
\begin{equation}
u=6\bar{u}\left(\frac{n}{t}-\frac{n^2}{t^2}\right),
\label{eq:paraprofile}
\end{equation}
where we introduced the thickness-averaged velocity 
\begin{equation}
\bar u = -\frac{t^2 \partial_s P_\mathrm{v}}{12\eta_\mathrm{v}}.
\end{equation}
The lubrication problem is closed using the axisymmetric continuity equation,
\begin{equation}
r\dot{t}+\partial_s\left(rt\bar{u}\right)=r  j,
\label{eq:cont}
\end{equation}
where $r$ is the distance from the symmetry axis and $\dot{t}$ is a time-derivative; in the remainder we will look for (quasi-)steady states (the evaporation time being much greater than the relaxation times) so that time derivatives can be omitted. The source term $j$ on the right hand side of (\ref{eq:cont}) is due to the flux of vapour generated by evaporation, which modelled by Fourier's law can be expressed as \citep{Maquet2016}
\begin{equation}
\label{eq:fourierlaw}
j=\frac{\epsilon}{t} \quad \mathrm{with} \quad \epsilon = \frac{k_\mathrm{v} \Delta T}{L\rho_\mathrm{v}}. 
\end{equation}
In this expression $k_\mathrm{v}$ and $\rho_\mathrm{v}$ respectively are the vapour thermal conductivity and density, $L$ the latent heat of evaporation and $\Delta T$ the temperature difference between the pool and the drop (the superheat).  Note that in the case of  boules of Hickman, the pool is superheated and evaporating, which can be modelled using Newton's law of cooling: $j=\mathpzc{h} \Delta T /(L\rho_\mathrm{v})$, where $\mathpzc{h}$ is the heat transfer coefficient and $\Delta T$ is based on the (superheated) pool temperature far away from the drop. In this case, $j$ is approximately constant along the film. For now, we focus on the Leidenfrost case of equation \ref{eq:fourierlaw}. The consequences of this different mechanism of vapour generation will be discussed in (\S\ref{sec:Scalinghickman}).

Thus, the vapour film is described by 
\begin{equation}
-\frac{1}{12\eta_\mathrm{v}\,r} \partial_s\left(rt^3 \partial_s P_\mathrm{v}\right)
%&=&
=
\frac{ \epsilon}{t} 
\label{eq:general1} 
%\\
%P_v&=& k-\rho_\mathrm{d} g h-\gamma_\mathrm{d} \kappa_h, \label{eq:general2} \\
%P_v&=&-\rho_\mathrm{p} g e+\gamma_\mathrm{p} \kappa_e, \label{eq:general3}
\end{equation}
in conjunction with equations (\ref{eq:pvpool}) and (\ref{eq:pvdrop}), which are three coupled equations for the vapour pressure $P_\mathrm{v}$ and respectively for the droplet and pool surface profiles $h$ and $e$. These equations need to be complemented by geometric expressions (see Appendix~\ref{app:numerics_full} for more details) for the interface curvatures $\kappa_{e,h}$, for $r$ as a function of $s$, viz.~$(\partial_s e)^2+(\partial_s r)^2=1$, and for the thickness of the vapour layer, viz.
\begin{equation}
t(s) \,\partial_s r= h(s)-e(s) .
\end{equation}

At the exit from the vapour film, where it joins the ambient atmosphere and where its thickness $t$ asymptotically diverges, the drop and pool surfaces are expected to attain equilibrium static shapes, described by 
\begin{equation}
P_\mathrm{v}=0
\label{eq:staticshapes}
\end{equation} 
with (\ref{eq:pvdrop}) and (\ref{eq:pvpool}), respectively. It may be useful to regard (\ref{eq:staticshapes}) as a degenerate form of (\ref{eq:general1}) as $t\to\infty$. It is matching with such static shapes that is imposed in one way or another (to be specified at each concrete occurrence) as the boundary conditions there. 

As for other boundary conditions, no singularity at the symmetry axis (i.e.~ at $r=0$) is imposed. At last, the pool surface attains its unperturbed level $e=0$ far away from the drop (as $r\to\infty$). 

The constant parameter $k$ entering the problem by means of (\ref{eq:pvdrop}) 
%can be represented for convenience as 
%\begin{equation}
%k=\rho_\mathrm{d} g h_\text{top}+\gamma_\mathrm{d} \kappa_{d,\text{top}} \,,
%\label{eq:kkk}
%\end{equation}
%where the subscript ``top'' refers to the values at the top of the drop. Clearly, the parameter $\kappa_{d,\text{top}}$   
is eventually the one quantifying the size of the droplet, and in this sense could be taken as one of the system parameters in the analysis (along with the material properties of the liquids and the superheat). However, we shall rather prefer to quantify the size more directly by means of the radius of the droplet's vertical projection, $R$. Hence, with $R$ as a system parameter, $k$ now becomes yet another unknown to be determined. 
%As for the parameter $h_\text{top}$, it fixes the vertical position of the upper, equilibrium part of the drop, and is un unknown of the problem to be determined from matching of the upper equilibrium shape with the vapour layer profile. 

For the time being, we shall keep the formulation in dimensional form. Appropriate non-dimensionalisations will rather be introduced later on in a context-specific way. However, one can already establish that the results are ultimately governed by four dimensionless parameters, which can be chosen as 
\begin{equation}
\Gamma=\frac{\gamma_\mathrm{p}}{\gamma_\mathrm{d}}, \quad \quad \mathcal{P} = \frac{\rho_\mathrm{p}}{\rho_\mathrm{d}}, \quad \quad \mathcal{R}=\frac{R}{\lambda_\mathrm{c}}, \quad \quad \tilde{\mathcal{E}} =
\frac{\eta_\mathrm{v} \epsilon }{\rho_\mathrm{d} g \lambda_\mathrm{c}^3}.
\label{numbers}
\end{equation}

The first two are the ratios of surface tension and density of the pool and the liquid. The third parameter $\mathcal{R}$ is the dimensionless radius of the drop, scaled by the capillary length $\lambda_\mathrm{c}=\left(\frac{\gamma_\mathrm{d}}{\rho_\mathrm{d} g}\right)^{1/2}$. Finally, the evaporation-induced viscous vapour flow is quantified by the evaporation number $\tilde{\mathcal{E}}$, proportional to the value of the superheat. Note that a key dimensionless number $\tilde{\mathcal{E}}$ naturally appears upon substituting (\ref{eq:pvdrop}) into (\ref{eq:general1}) and normalizing all the length variables with $\lambda_\mathrm{c}$ \citep{Sobac2014,Maquet2016}. 

%\\
%Calculations for the superhydrophobic drop are done by  solving the Young-Laplace equations for both the drop and the pool, subject to differing liquid properties. The interfacial tension between the pool and the drop $\gamma_{dp}$ is the sum of both the pool and the drop, resulting in a contact angle of \SI{180}{\degree}.

\subsection{Asymptotic approach}
\label{sec:asy}

The values of the evaporation number $\tilde{\mathcal{E}}$ encountered in practice are typically rather small \citep{Snoeijer2009,Celestini2012, Sobac2014,Maquet2016}. This is what eventually justifies the very structure of the problem assumed in \S\ref{sec:model}: a thin vapour layer between the substrate (here the pool), where the lubrication approximation is applicable, and equilibrium shapes of liquid surfaces (drop and pool) beyond the thin vapour layer. Direct computations of the Leidenfrost problem carried out under this premise confirm its self-consistency for both flat solid substrates \citep{Snoeijer2009,Sobac2014} and deformed liquid ones \citep{Maquet2016}. Essential deviations from this scheme are only encountered, on the one hand, for sufficiently small drops, well below the capillary length \citep{Celestini2012,Sobac2014}. In the present paper, we shall deal only with drops well larger than that anyway (see below), and hence with no such limitation. On the other hand, the scheme breaks down on the verge of the chimney instability \citep{Snoeijer2009,Sobac2014}, which, as already anticipated, will not be encountered in the present case of a liquid substrate. 

Thus, in fact, the formulation provided in \S\ref{sec:model} already tacitly implies $\tilde{\mathcal{E}}\ll 1$. Direct numerical computation for such a ``full'' formulation can be realized e.g.\ using an approach largely similar to \cite{Sobac2014} and \cite{Maquet2016}. Here it is just rendered geometrically more elaborate in order to handle large deformations of the pool surface, see Appendix~\ref{app:numerics_full} for more details. 

However, the fact that $\tilde{\mathcal{E}}\ll 1$ also opens the way to a systematic asymptotic analysis, consistently and thoroughly exploiting this limit. This is actually what the present paper is about. As it is typically the case, such an asymptotic analysis will permit further insight into the physics of the problem. Here we shall in particular be interested in further details as far as the structure of the vapour layer is concerned. Importantly, this will also permit establishing the scaling with $\tilde{\mathcal{E}}$ of various quantities of interest. In the case of a flat solid substrate, such a program has been realized e.g.\ by \cite{Snoeijer2009} and \cite{Sobac2014}. As already said, we anticipate essential differences in the case of a liquid substrate we are concerned with in the present paper. The hereby obtained asymptotic results will be validated against the earlier mentioned direct numerical simulation in the framework of the full formulation of \S\ref{sec:model}. 

With the expected asymptotically small vapour-layer thickness $t$ in the limit of small evaporation numbers $\tilde{\mathcal{E}}\ll 1$, one actually realizes that the shape of our Leidenfrost drop must be asymptotically close to that of an equilibrium ``superhydrophobic'' drop. The shape of the latter is governed by the static balance between surface tension and gravity. Namely, in this superhydrophobic configuration, the liquid--liquid interface is determined by (\ref{eq:pvpool}) and (\ref{eq:pvdrop}) with $e\equiv h$ and formally possesses an interfacial tension $\gamma_{dp}\equiv \gamma_\mathrm{p}+\gamma_\mathrm{d}$; the liquid--gas interface is described by (\ref{eq:pvdrop}) with (\ref{eq:staticshapes}) for the drop itself, and by (\ref{eq:pvpool}) with (\ref{eq:staticshapes}) for the pool; at the contact (triple) line, the slopes of all the three interfaces coincide, and a contact angle of $180^\circ$ results from the side of each of the liquids (see Appendix~\ref{app:superhydrophobic} for the computation method). It is important to note that the picture in terms of a superhydrophobic drop refers to a large-scale description: in the actual non-equilibrium configuration, the ``contact line" in fact represents the position of a thin neck region through which the vapour escapes to the surrounding atmosphere. 

Once the superhydrophobic shape is known, the leading-order pressure distribution governing the vapour flow in (\ref{eq:general1}) can immediately be drawn from (\ref{eq:pvpool}), or what is the same from (\ref{eq:pvdrop}). Quite remarkably, this distribution will clearly not be constant in the pool case, with an appreciably non-flat surface, i.e.\ we end up with $\partial_s P_v\ne 0$ independently (to leading order) from the actual vapour layer profile $t(s)$. This is in stark contrast with a drop on a flat rigid substrate, for which $\partial_s P_v\ne 0$ {\it does} depend upon $t(s)$ (i.e.\ upon a higher-order approximation in terms of $\tilde{\mathcal{E}}\ll 1$). We shall see that this is the key factor giving rise to different vapour-layer structures and scalings for the two types of substrate. 

However, a common feature between the two types of substrate is that $P_v$ will not be continuous across the ``triple line", i.e. across the thin neck region through which the vapour escapes from below the drop. This is due to a mismatch in curvature on both sides of the neck, where $P_v$ exhibits a jump from $P_v>0$ under the drop to the ambient value $P_v=0$. In experiments however $P_v$ will be continues, which will be discussed in more detail in \autoref{sec:mismatch}. The appearance of this thin neck allows us to introduce and exploit the following hierarchy of length scales

\begin{equation}\label{eq:hierarchy_main}
t_\mathrm{n} \ll t_0 \ll \lambda_\mathrm{c} \sim R\,,
\end{equation}
where $t_0$ and $t_\mathrm{n}$ are, respectively, the typical vapour layer thicknesses outside and inside the neck region. Such a presence of separate distinguished regions with different spatial scales makes us resort to matched asymptotic expansions as the most appropriate asymptotic method for the problem at hand. Overall, it turns out that the problem can be split into the following regions, indicated schematically in figure~\ref{fig:sketchzones}. At large scales, there are two outer regions: 1) below the drop, and 2) above the drop and pool. In the limit of vanishing $t$, these outer solutions will precisely correspond to superhydrophobic drops on a liquid pool. As already mentioned, these are equilibrium solutions that can be computed from the static balance between surface tension and gravity. These two outer regions are connected by a smaller inner region, which is nothing else than the earlier mentioned neck region. It will turn out that no direct matching between the outer and inner regions is possible in the vapour layer, and so yet another, intermediate region, also marked in figure~\ref{fig:sketchzones}, will have to be implied.  

As already pointed out, the here expected completely new type of solution and $\tilde{\mathcal{E}}$-scalings for the vapour layer basically owe themselves to the substrate surface deformability. Note though that for smaller drops, $R<\lambda_\mathrm{c}$ ($\mathcal{R}<1$), the pool surface gets increasingly more flat \citep{Maquet2016}. Therefore, for sufficiently small drops, the appropriate asymptotic theory is likely to involve other smallness parameters apart from $\tilde{\mathcal{E}}$, viz.\ a geometric one characterizing the small substrate non-flatness. Another kind of limitation for smaller drops and related to the mentioned intermediate region will be pointed out in \S\ref{sec:finite}. However, in essence, we shall leave smaller drops beyond the scope of the present paper by rather focusing on sufficiently large ones. 

Furthermore, a large part of our analysis (\S\ref{sec:analysis}) will be dedicated to really large drops, $R\gg \lambda_\mathrm{c}$ ($\mathcal{R}\gg 1$).  The expected hierarchy of length scales (\ref{eq:hierarchy_main}) then rewrites as 
\begin{equation}\label{eq:hierarchy}
t_\mathrm{n} \ll t_0 \ll \lambda_\mathrm{c} \ll R \,.
\end{equation}
Besides, inspired by the boules of \citet{Hickman1964}, we launch the analysis by considering liquids with equal properties, $\Gamma=1$ and $P=1$. Mathematically, this gives rise to the simplest possible configuration (the associated superhydrophobic drop assuming a hemispherical shape), where analytical solutions are possible and which is a good starting point for developing the essence of our asymptotic approach. Subsequently, in \S\ref{sec:extend}, we extend our analysis to smaller drops, $R\sim \lambda_\mathrm{c}$ ($\mathcal{R}\sim 1$), and to non-equal liquid properties, $\Gamma\ne 1$ and/or $\mathcal{P}\ne 1$.

\section{A large drop on a pool with the same mechanical properties}\label{sec:analysis}
Here we perform a detailed analysis of a situation resembling the ``boules" of \citet{Hickman64}, shown in \autoref{fig:HickBoule}. In this case the pool and the drop  have equal density and surface tension, i.e. $\Gamma=\mathcal{P}=1$. The boules formed are much larger than the capillary length, $\mathcal{R} \gg 1$, and below we will consider the asymptotics for small evaporation numbers, $\mathcal{E} \ll 1$. 

\subsection{Outer region 1: below the drop}\label{sec:q}

\subsubsection{The droplet shape.}
The outer shape of the drop and pool can be obtained by combining (\ref{eq:pvpool}) and (\ref{eq:pvdrop}):
\begin{equation}\label{eq:bla}
0=k - \rho g(h-e) - \gamma ( \kappa_h+\kappa_e),
\end{equation}
where we assumed identical material properties $\gamma=\gamma_\mathrm{d}=\gamma_\mathrm{p}$ and $\rho=\rho_\mathrm{d}=\rho_\mathrm{p}$. We anticipate the gap thickness, $t \sim h-e$, to be much smaller than the radius, in which case $\kappa_h \simeq \kappa_e$. The hydrostatic term can also be neglected when $t\sim(h-e)\ll 4\gamma/\rho g R \sim \lambda_c^2/R$, a condition that is much more severe and will be monitored for our solution \textit{a posteriori}. Equation (\ref{eq:bla}) then further simplifies, 
\begin{equation}\label{eq:bla1}
0= k - 2 \gamma \kappa_h,
\end{equation}
imposing a shape of constant curvature. Below we will find that the matching condition requires the outer solution to be a perfect hemisphere. Hence, we can identify $k=4\gamma/R$, where $R$ is the maximum radius of the drop. 

\subsubsection{The vapour thickness.}

\begin{figure}
\centering
\includegraphics[width=\textwidth]{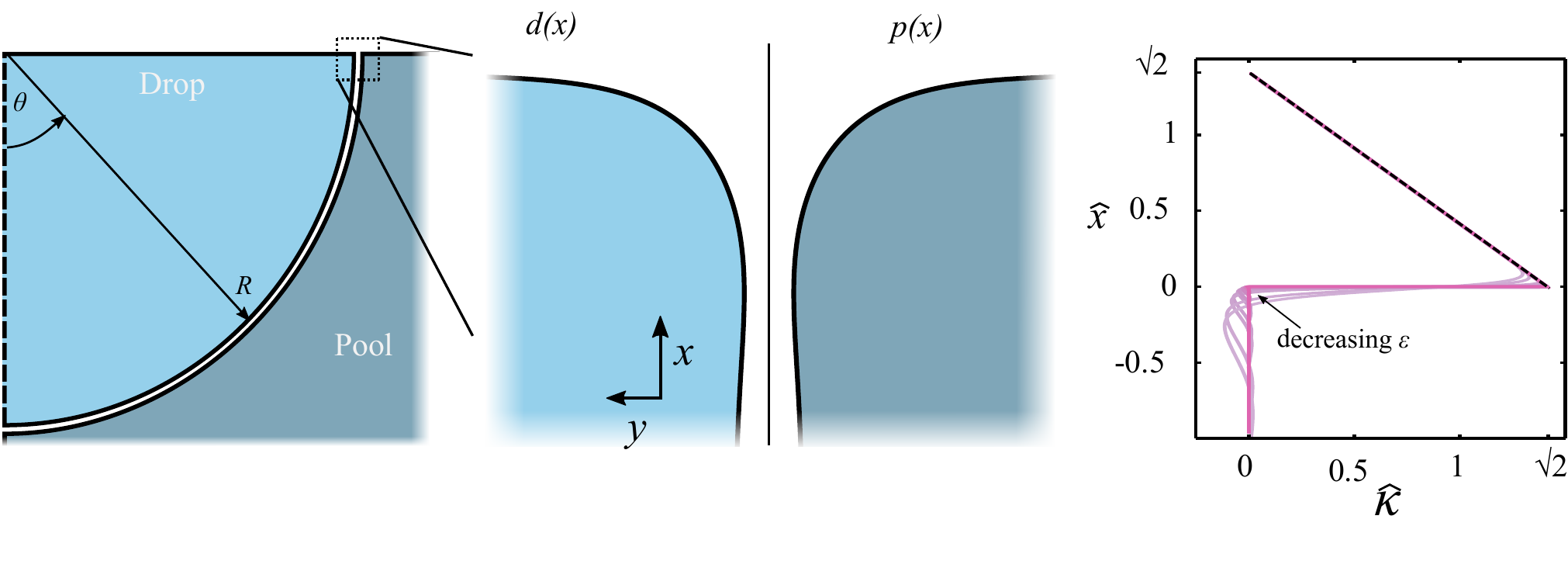}
\caption{Left: Sketch of outer region 1. Since the shape below the drop reduces to a hemisphere, we adopt a spherical coordinate system where the gap thickness $t$ depends on the angle $\theta$. The middle panel shows the local Cartesian coordinate system around $\theta=\pi/2$, which is used for the analysis of the thin neck region at the exit of the vapour layer. The drop and pool interfaces are expressed as $d(x)$ and $p(x)$ respectively, while the vapour thickness reads $t=d-p$. Right: The interface curvature $\kappa$ in the neck region, obtained numerically from (\ref{eq:neck}) for various evaporation rates $\mathcal{E}$. A distinctive jump in curvature can be seen, indicating a sharp pressure jump through the thin neck. The dashed line shows the puddle solution for outer region 2.
}
\label{fig:SketchSphere}
\end{figure}

The spherical geometry of the outer solution, and thus of the vapour layer, suggests that the analysis of the lubrication flow will be most easily expressed using spherical coordinates (see \autoref{fig:SketchSphere}). In this coordinate system, the lubrication equation (\ref{eq:general1}) reads
\begin{eqnarray}\label{eq:wafwaf}
-\frac{1}{12\eta_\mathrm{v} R^2\ \sin{\theta}}\partial_\theta \left[\sin{\theta}\ t^3\ \partial_\theta P_v\right]&=&\frac{\epsilon}{t}.
\end{eqnarray}
This equation needs to be complemented by an equation for the pressure gradient $P'_v$, for example (\ref{eq:pvdrop}), which in spherical coordinates simplifies to $\partial_\theta P_v = -\rho g R\sin \theta$. The lubrication equation can then be expressed as
\begin{equation}\label{eq:ploep}
\frac{1}{2 t^3 \sin{\theta}}\partial_\theta \left[t^3 \sin^2{\theta} \right]
= \frac{6 \eta_\mathrm{v} \epsilon R}{\rho g t^4}.
\end{equation}
In this expression we collected all dimensional parameters on the right hand side, to form a dimensionless ratio.

The above expression invites us to introduce a dimensionless thickness $\tilde t = t/t_*$, where the characteristic scale for the thickness reads

\begin{equation}\label{eq:t0}
t_*=\left( \frac{6 \eta_\mathrm{v} R \epsilon}{\rho g} \right)^{1/4} = \lambda_\mathrm{c} \, \mathcal{E}^{1/4},
\end{equation}
where a (modified) evaporation number 
$$\mathcal{E}= \frac{6\eta_\mathrm{v} R \epsilon}{\rho_\mathrm{d} g \lambda_\mathrm{c}^4} =\frac{6 R}{\lambda_\mathrm{c}}\,\tilde{\mathcal{E}}$$
has been introduced, as it naturally occurs in this form in the present $\mathcal{R}\gg 1$ context.
Inserting this rescaling in (\ref{eq:ploep}) and working out the derivatives gives
\begin{eqnarray}
\tilde{t}^3 \partial_\theta\tilde{t} &=&\frac{2\left[1 -\tilde{t}^4\ \cos{\theta}\right]}{3 \sin{\theta}\ },
 \label{eq:sphere}
\end{eqnarray}
which is a first order ODE for the profile of the vapour layer $\tilde t(\theta)$. The solution to this equation can be cast in closed form
\begin{eqnarray}
\label{eq:analytical}
\hspace{-0.2in}\left[\tilde t(\theta)\right]^4 &=&  \frac{8 \int_0^\theta dx \, (\sin x)^{5/3 } }{3(\sin \theta)^{8/3}} \nonumber\\
&=& \frac{8}{15} \frac{1}{(\sin{\theta})^{8/3}} \left(\frac{\pi^\frac{1}{2} \Gamma  \left( \frac{1}{3} \right)}{\Gamma  \left( \frac{5}{6} \right)} - 
   \cos{\theta} \left(2\; {}_2 \text{F}_1 \left[\frac{1}{2}, \frac{2}{3}, \frac{3}{2},( \cos{\theta})^2\right] +       3 (\sin{\theta})^\frac{2}{3}\right)\right).
\end{eqnarray}
Here  we imposed a non-singular behaviour at the symmetry axis, only possible with $\tilde t_0\equiv \tilde t(0)=1$, which was anticipated in the definition in (\ref{eq:t0}). Here  $\Gamma(x)$ is the Gamma function, ${}_2\text{F}_1$ is a generalized hypergeometric function \citep{MathWorld_HF} and this expression was obtained using Mathematica. When plotting the solution (\ref{eq:analytical}), one observes that the vapour thickness is nearly constant. It very mildly increases from $\tilde t=1$ at $\theta=0$, to a slightly larger value at $\theta=\pi/2$ (the exit of the gap):

\begin{equation}\label{eq:magic}
\tilde t_{\rm exit} = 2^\frac{3}{4} 15^{-\frac{1}{4}} \pi^\frac{1}{8} \left( \frac{\Gamma  \left( \frac{1}{3} \right)}{\Gamma  \left( \frac{5}{6} \right)}\right)^\frac{1}{4} \approx 1.22386\cdots\,.
\end{equation}

\subsubsection{Summary and comparison to numerical solution.}

\begin{figure}
\includegraphics[scale=1]{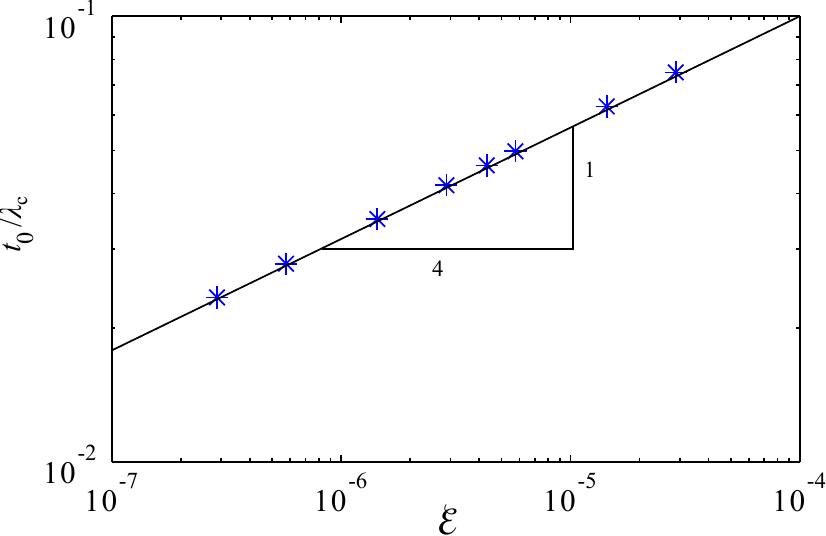}
\centering
\caption{Results for the central thickness $t_0\equiv t(0)$ of the vapour layer \textit{versus} the evaporation number for a drop of $R= 10 \lambda_\mathrm{c}$ on a pool having the same properties as the drop. The data points (numerical solution to the full problem) approach the curve $t_0=\lambda_\mathrm{c} \mathcal{E}^{1/4}$ obtained by asymptotic methods in the limit of small $\mathcal{E}$.}
\label{fig:t0E}

\end{figure}

An important set of analytical results has thus been obtained. First, we found that the immersed part of the drop takes a hemispherical shape. Second, the vapour layer thickness at the axis below the drop is given by $\tilde t_0=1$. Third, the profile of the vapour layer in the outer zone is characterised by a single, universal profile, given by the expression (\ref{eq:analytical}). For the matching to the neck region (see inset figure~\ref{fig:SketchSphere}), an important result is the thickness at the ``gap exit" at $\theta = \pi/2$, for which the analytical expression (\ref{eq:magic}) is found. In dimensional form, we thus obtain the relevant thicknesses
\begin{equation}
t_0 = \lambda_\mathrm{c} \mathcal{E}^{1/4}, \quad \quad
t_{\text{exit}} = 1.22386\cdots\, \lambda_\mathrm{c} \mathcal{E}^{1/4}. 
\label{eq:t0_texit_dim}
\end{equation}
These results were obtained under the assumption of the hierarchy of scales (\ref{eq:hierarchy}), which is self-consistent as long as
\begin{equation}
\mathcal{E}^{1/4} \ll 1 \ll \mathcal{R} \ll \mathcal{E}^{-1/4}	.
\label{eq:Epsordering}
\end{equation}
Note the latter, strong inequality arising form neglecting the hydrostatic pressure difference across the vapour gap $(t\ll \lambda_\mathrm{c}^2/R$). To validate these findings, we compared the results to numerical solutions of the full problem formulated in \S\ref{sec:model} (see also Appendix~\ref{app:numerics_full}). Indeed, it is observed that for large drop volumes the immersed part of the droplet approaches a perfect hemisphere. In \autoref{fig:t0E} we provide a further quantitative test, by plotting the central thickness $t_0$ for a droplet of radius $R=10\lambda_\mathrm{c}$. Upon reducing the evaporation number $\mathcal{E}$, the numerical result perfectly approaches the prediction (\ref{eq:t0_texit_dim}). Moreover,  note that the last inequality (\ref{eq:Epsordering}) is well satisfied in the given range of $\mathcal{E}$. Evaluating $t_0$ for the experimental conditions of \autoref{fig:HickBoule}, we obtain $t_0\approx \SI{70}{\micro\meter}$.

\subsection{Outer region 2}\label{sec:outer2}

\subsubsection{Puddle solutions.}

We now turn to the second outer region, describing the top of the droplet and of the pool. In this range the vapour is no longer confined to a thin gap, so that viscous effects are completely negligible here. By consequence, the pool and drop are described by (\ref{eq:pvpool}) and (\ref{eq:pvdrop}) with $P_v=0$. We further note that the constant $k \rightarrow 0$ for large drops, so that the profiles of the drop and the pool become each others mirror images. The solutions that result from these equations are the classical puddle solutions, for which the curvature increases linearly with depth. For explicit forms we refer to \citet{deGennes2013, landau1959lifshitz}.

\subsubsection{Mismatch with outer solution 1.}
\label{sec:mismatch}
Importantly, the puddle solution exhibits a finite curvature at the droplet's edge, namely $\kappa= 2^{1/2}/\lambda_\mathrm{c}$. This value of the curvature is obtained since, by symmetry, the puddle approaches the neck region vertically. This is to be contrasted with the curvature in the other outer region, below the drop, for which the curvature was found to be $1/R$. For large drops, we thus find a ``mismatch" in curvature near the exit of the vapour layer. This implies that the problem requires an inner zone that smoothly joins the two outer regions. The matching is illustrated in the right panel of  \autoref{fig:SketchSphere}, where we present a numerical solution to the problem (details are given in the paragraphs below). The centre panel provides a detailed view of the droplet and pool profiles, and reveals a thin neck that is connected to the outer vapour layer. The right panel shows the corresponding curvatures, for various values of $\mathcal{E}$. The dashed line corresponds to the puddle solution; the numerical profiles smoothly join the puddle to the vapour film of vanishing curvature.

\subsection{Inner region: the neck profile}
\label{sec:inner}

\subsubsection{Matching conditions and numerical solution}
\label{seq:reg2neck}
By inspection of \autoref{fig:SketchSphere}, the thin neck region represents a small vertical zone around $\theta=\pi/2$. We therefore adopt a local Cartesian coordinate system, as sketched in \autoref{fig:SketchSphere}, where gravity acts along the $x$-axis towards $x<0$. In this coordinate frame, we describe the drop interface as $d(x)$ and the pool interface as $p(x)$. The gap thickness is then expressed as $t(x)=d(x)-p(x)$ and its curvature $\kappa_t(x)=\kappa_\mathrm{d}(x)-\kappa_\mathrm{p}(x)$. The expressions for the pressure in the film from (\ref{eq:pvpool}) and (\ref{eq:pvdrop}), for equal liquid properties and $k=0$, can be written as
\begin{eqnarray}
P_v=-\rho g x-\gamma \kappa_\mathrm{d}\nonumber\\
P_v=-\rho g x+\gamma \kappa_\mathrm{p}\nonumber.
\end{eqnarray}
Taking the difference of these equations, one finds $\kappa_\mathrm{d}=-\kappa_\mathrm{p}=\frac{1}{2}\kappa_t$, resulting in a symmetric deformation for both $d$ and $p$. Equation (\ref{eq:general1}) then becomes

\begin{eqnarray}\label{eq:blaa}
\frac{1}{12\eta_\mathrm{v}} \partial_{{x}} \left[{t}^3  \partial_{{x}} \left[\rho g {x}+\frac{1}{2}\gamma \partial_{xx} t  \right] \right]&=&\frac{\epsilon}{{{t}}}\approx 0, 
\end{eqnarray}
where we used the small slope representation of the curvature $\kappa_t=t''$ for consistency with the lubrication approximation. Since we expect the local vapour generation in the small inner region to be negligible compared to the total generated flux, we can also drop the right hand side. Ultimately the gap profile needs to be matched to that of the puddle shape $d(x)$. It is therefore convenient to express the lubrication profile by $d(x)=t(x)/2$, and non-dimensionalize all lengths by $\lambda_\mathrm{c}$:
\begin{equation}
\hat{d} = \frac{d}{\lambda_\mathrm{c}} =\frac{t}{2\lambda_\mathrm{c}}, \quad \quad \hat{x}=\frac{x}{\lambda_\mathrm{c}}.
\end{equation}
With this, we obtain after integrating (\ref{eq:blaa}) once:
\begin{equation}
\hat{d}'''=\frac{c^3}{\hat{d}^3}-1,
\label{eq:neck}
\end{equation}
where $c$ is an integration constant. This equation describes the shape of the drop interface inside the neck region.

The problem is closed by the matching conditions to the outer regions 1 and 2, respectively corresponding to negative and positive limits of $\hat x$. The curvature of outer region 1 scales as $1/R$, which in dimensionless variables gives $\hat d'' \sim 1/\mathcal{R}$ and is thus asymptotically small for large drops, implying $\hat{d}'(-\infty)=\hat{d}''(-\infty)=0$. Therefore, we require $\hat{d}(-\infty)$ approaches a constant value, that according to (\ref{eq:neck}) can be equated to $c$. This thickness must ultimately match the ``exit" thickness of outer zone 1, leading to the condition

\begin{equation}\label{eq:matchgap}
\frac{t_{\rm exit}}{2} = d(-\infty) \quad \Longrightarrow \quad c = \frac{1}{2} \cdot 1.22386\cdots \mathcal{E}^{1/4}. 
\end{equation}
The boundary condition for positive $\hat x$ comes from matching the curvature of the puddle solution, which in scaled units  reads $\hat d'' = \sqrt{2}$ (see $\S$\ref{sec:outer2}). We numerically solved (\ref{eq:neck}) subject to these boundary conditions. 

\begin{figure}
\centering
\includegraphics[width=0.95\linewidth]{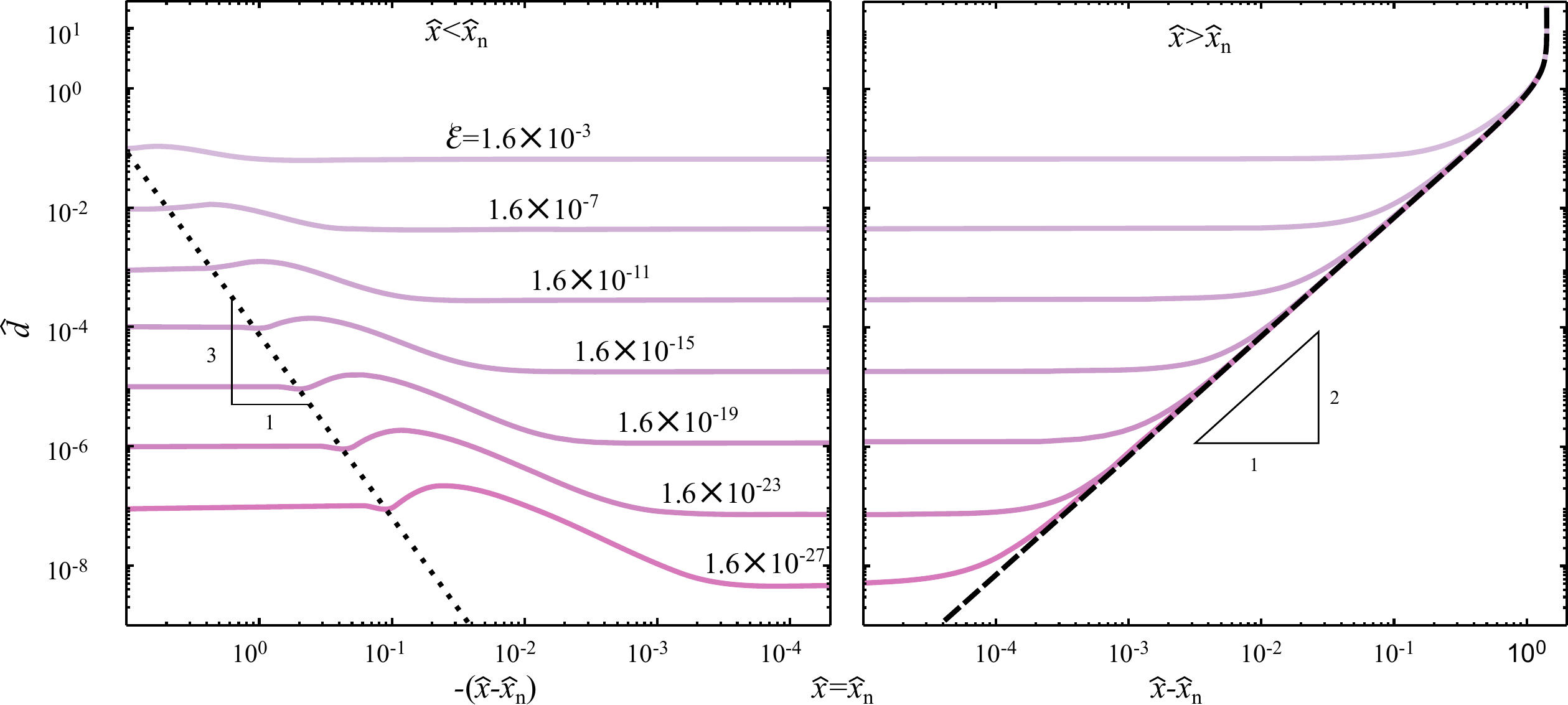}
\caption{Neck profiles obtained from numerical solution of (\ref{eq:neck}) for various vapour layer thicknesses $\hat{d}(-\infty) \sim \mathcal{E}^{1/4}$ for the case of equal drop and pool properties and large drops ($R\gg\lambda_\mathrm{c}$). The double logarithmic representation on the two panels, with inverse log-scale centered around the neck position $\hat{x}=\hat{x}_\mathrm{n}$, reveals the details of the thin neck region and the oscillations upon approaching the vapour film. The dotted line indicates the location of the oscillations, scaling as $\hat{d}\sim (\hat x_\mathrm{n}-\hat x)^3$. The dashed line shows the puddle solution for outer region 2, exhibiting a $\hat{d}\sim (\hat x-\hat x_\mathrm{n})^2$ upon approaching the neck.}
\label{fig:Curvesneck}
\end{figure}

The resulting neck profiles are presented \autoref{fig:Curvesneck} for different values of $\mathcal{E}$, corresponding to different gap thickness according to (\ref{eq:matchgap}). One can observe that reducing the flux $\mathcal{E}$ leads to a localisation of the neck region, both in terms of thickness and lateral extent. To highlight these trends we have reported the profiles on two panels of double logarithmic scales, centred around the position $\hat x_\mathrm{n}$ of the minimum neck thickness (see caption for details). The same data were used to show the mismatch in curvature in \autoref{fig:SketchSphere} (right panel). It is interesting to note that the profiles are strongly reminiscent of dimple profiles observed in dip-coating \citep{Snoeijer2008}, which are indeed governed by an equation similar to (\ref{eq:neck}). Apart from the dimple, the dip-coating solutions also exhibit the oscillations seen in the left panel of \autoref{fig:Curvesneck}, which were analysed in detail by \citet{Benilov2010}.

\subsubsection{Self-similar solution for the neck region}
\label{sec:selfsim}
Based on our numerical results we observe that the neck region near position $\hat x_\mathrm{n}$ becomes increasingly localized for small $\mathcal{E}$, while the neck thickness $\hat d_\mathrm{n}$ is found to decrease. This is in direct analogy to the neck region for normal Leidenfrost drops, above a rigid surface \citep{Snoeijer2009,Sobac2014}. Owing to the smallness of $\hat d$ inside this region, (\ref{eq:neck}) reduces to
\begin{equation}\label{eq:neck2}
\hat{d}^3\hat{d}'''=c^3,
\end{equation}
which means that the gravity is subdominant with respect to viscosity and surface tension. Indeed, (\ref{eq:neck2}) is  identical to the neck equation studied by \citet{Snoeijer2009}, which admits similarity solutions
\begin{equation}\label{eq:ansatz}
\hat{d}(x)=c^\alpha \, T(\zeta),\ \ \textrm{with} \ \ \zeta=\frac{\hat x-\hat x_\mathrm{n}}{c^\beta}.
\end{equation}
Inserting this Ansatz in (\ref{eq:neck2}) gives
\begin{equation}
T^3T'''=1, \ \ \textrm{and}\ \  4\alpha-3\beta=3 .
\label{eq:selfsim}
\end{equation}
The exponents $\alpha,\beta$ can be determined from a matching condition for $\zeta \gg 1$, for which the shape of the pool $\hat d(x) \simeq (\hat x-\hat x_\mathrm{n})^2/\sqrt{2}$ must be approached, regardless of the value of $c$. This implies for large $\zeta$,
\begin{equation}
c^\alpha \,  T(\zeta) \simeq c^\alpha \,  \zeta^2/\sqrt{2} = c^{\alpha-2\beta} \, (\hat x-\hat x_\mathrm{n})^2/\sqrt{2},
\end{equation}
hence, $\alpha-2\beta=0$. Combined with (\ref{eq:selfsim}), this gives the exponents $\alpha=6/5$ and $\beta=3/5$.

While for large positive $\zeta$ we can impose the asymptotic boundary condition $T'' \simeq \sqrt{2} $, we still need to provide the asymptotics for negative $\zeta$. We now show that the matching to the film region implies $T''\simeq 0$ as the missing boundary condition. The matching to the film can in principle be obtained from a detailed analysis of the oscillatory approach to the thin film, in the spirit of the work on dip-coating \citep{Benilov2010}. Here we focus only on the first oscillation, which is sufficient for the present purpose. The typical slope of the neck solution $\hat d' \sim c^{3/5} \sim \mathcal{E}^{3/20}$ must be compared to that of the first bump. This bump has its own thickness scale $\hat{\delta}_\mathrm{b}$ and lateral scale $\hat{\ell}_\mathrm{b}$, such that we demand $\hat{\delta}_\mathrm{b}/\hat{\ell}_\mathrm{b} \sim  \mathcal{E}^{3/20}$. For the approach of the bump we argue that all terms in (\ref{eq:neck}) are involved, such that $\hat d'''$ must be of order unity, or $\hat{\delta}_\mathrm{b} /\hat{\ell}_\mathrm{b}^3 \sim \mathcal{E}^0$. This scaling indeed gives the correct estimation for the position of the first oscillations in the left panel of \autoref{fig:Curvesneck} (dotted line). Combining these two equations on $\hat{\delta}_\mathrm{b}$ and $\hat{\ell}_\mathrm{b}$ we find $\hat{\delta}_\mathrm{b} \sim \mathcal{E}^{9/40}$ and $\hat{\ell}_\mathrm{b} \sim \mathcal{E}^{3/40}$. The final step is to evaluate the curvature of the bump $\hat{\delta}_\mathrm{b} /\hat{\ell}_\mathrm{b}^2 \sim  \mathcal{E}^{3/40}$, which as anticipated, vanishes for small $\mathcal E$.

\begin{figure}
\includegraphics[scale=0.5]{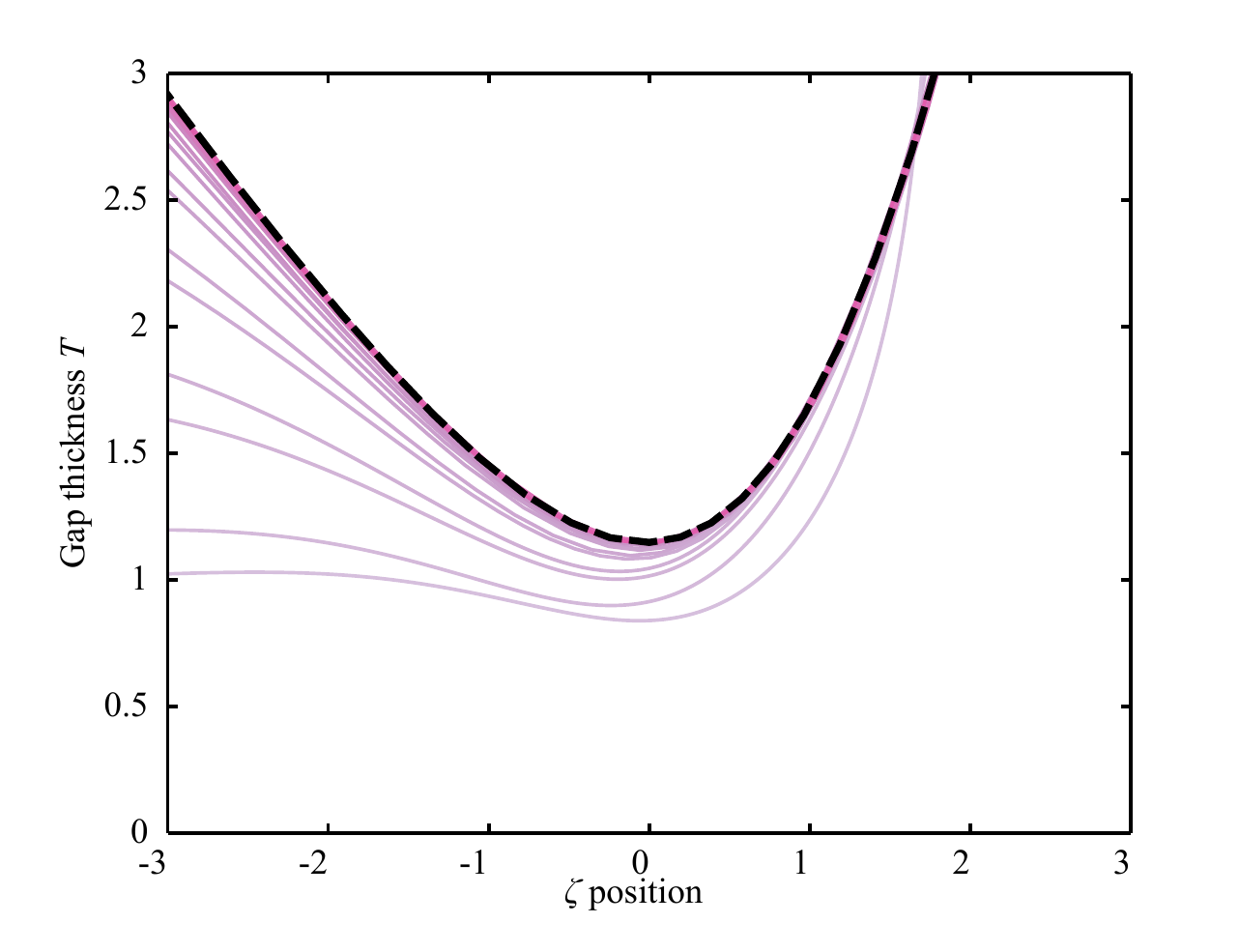}
\includegraphics[scale=0.50]{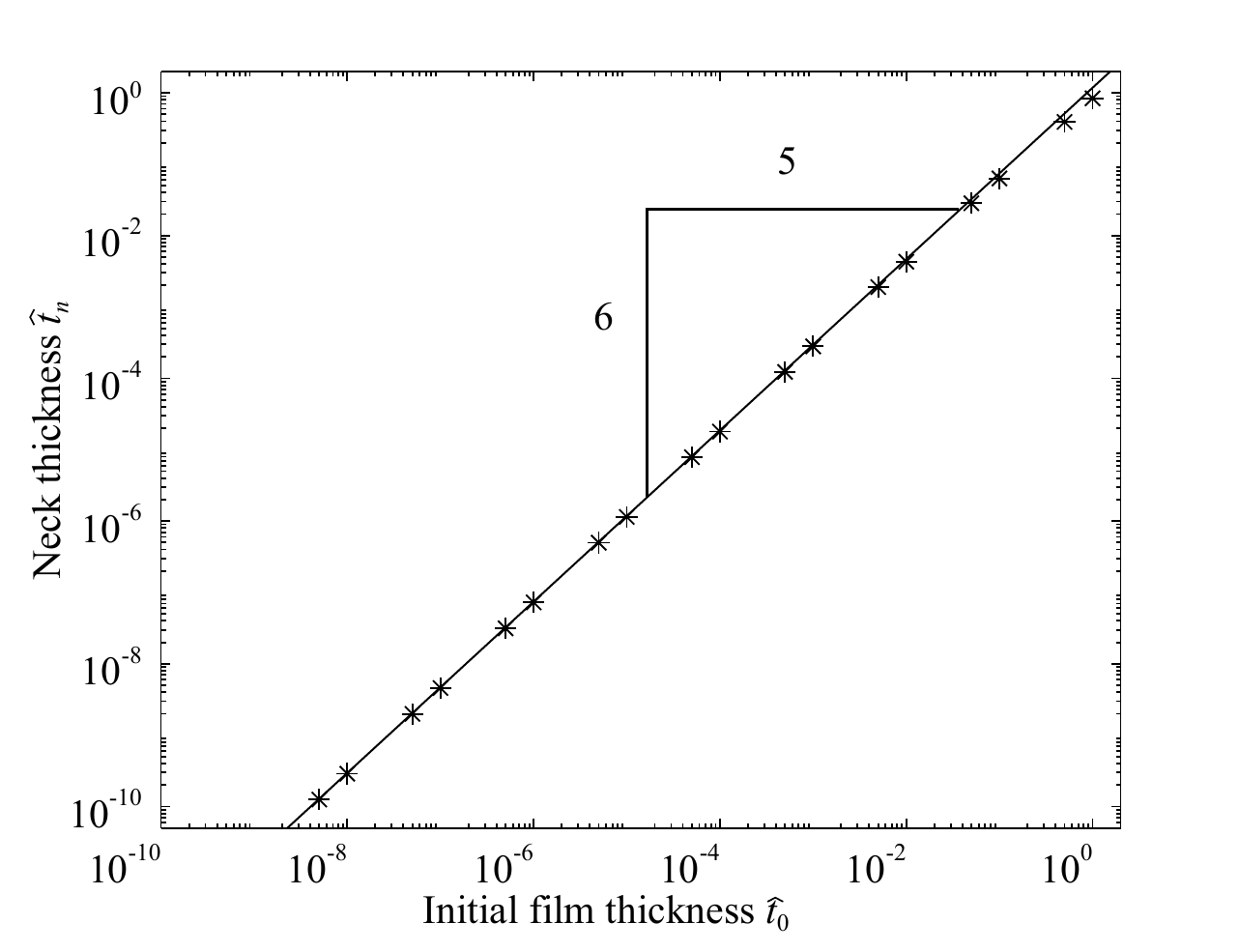} 
\caption{The neck region exhibits a self-similar structure, captured by the similarity function $T(\zeta)$ given by the dashed line (left panel). The other curves represent numerical profiles of the neck shown in \autoref{fig:Curvesneck}, scaled according to $T = \hat d/c^{6/5}$ and $\zeta = (\hat x- \hat x_\mathrm{n})/c^{3/5}$. Each of these profiles corresponds to a dot in the right panel, where the straight solid line corresponds to the similarity law (\ref{eq:tn}). The case of equal drop and pool properties and large drops ($R\gg\lambda_\mathrm{c}$). } 
\label{fig:Minneck}
\end{figure}

In summary, we expect the neck to be governed by a similarity solution $T(\zeta)$, which can be computed from (\ref{eq:selfsim}) subject to boundary conditions $T''(-\infty) =0$ and $T''(\infty)=\sqrt{2}$. The numerical solution is given in \autoref{fig:Minneck}, represented as a dashed line. The other curves correspond to the profiles of \autoref{fig:Curvesneck},  scaled according to (\ref{eq:ansatz}). We observe a collapse onto the similarity solution as the value of $c \sim \mathcal{E}^{1/4}$ is reduced. The relation for the minimum neck thickness can now be found by determining the minimum of the similarity function, which we numerically find to be $T_\mathrm{n}=1.147 \cdots$. Hence, we find 
\begin{equation}\label{eq:tn}
\hat d_\mathrm{n} = 1.147\cdots c^{6/5},
\end{equation}
which provides the minimum thickness at the neck, as confirmed in the right panel of \autoref{fig:Minneck}.

%
%
%The proper rescaling then becomes:
%
%\begin{equation}
%\hat{t}=\hat{t}_{-\infty}^\frac{6}{5}T\left(\frac{x-x_\mathrm{n}}{\hat{t}_{-\infty}^\frac{3}{5}}\right).
%\label{eq:rescale}
%\end{equation} 
%Evaluating the neck thickness as a function of initial thickness $\hat{t}_{-\infty}$, indeed reveals a 6/5 - scaling law, as shown in the right panel of \autoref{fig:Minneck}.
%When the profiles of \autoref{fig:Curvesneck} is rescaled in this way, a collapse is found for vanishing flux, shown in the inset. The solution for \autoref{eq:selfsim} is now found by  evaluating the asymptotic behaviour for $\zeta \rightarrow \pm\infty$. In the case of $\zeta \rightarrow -\infty$, the profile is to match the smooth film solution i.e. $T''=0$. For $\zeta \rightarrow+\infty$ the profile needs to match the puddle solution for outer region 2: $T''=\sqrt{2}$. Numerical evaluation of \autoref{eq:selfsim} is represented  by a dashed line in \autoref{fig:Curvesneck}, from which we obtain the minimal thickness $T_\mathrm{n}=1.147$. 

\subsection{Summary}

Let us now conclude the analysis for $\mathcal{E} \ll 1$, $\mathcal{R} \gg 1$ for the case where the drop and the pool consist of the same liquid ($\Gamma=1$, $\mathcal{P}=1$). We first recall the expressions for the vapour layer thickness below the center of the drop ($t_0$) and the vapour thickness as it approaches the neck ($t_{\rm exit}$):
\begin{equation}
t_0 = \lambda_\mathrm{c} \mathcal{E}^{1/4}, \quad \quad
t_{\text{exit}} = 1.22386\cdots \lambda_\mathrm{c} \mathcal{E}^{1/4}.  \nonumber
\end{equation}
These can now be complemented by the minimum thickness of the neck
\begin{equation}
t_\mathrm{n} = 2 \lambda_\mathrm{c} \hat d_\mathrm{n} = 1.272 \cdots  \lambda_\mathrm{c} \mathcal{E}^{3/10},
\end{equation}
which was obtained using (\ref{eq:matchgap}) and (\ref{eq:tn}). 
The hierarchy of scales (\ref{eq:hierarchy}) is indeed satisfied and the approach is self-consistent as long as
\begin{equation}
\mathcal{E}^{3/10} \ll \mathcal{E}^{1/4} \ll 1 \ll \mathcal{R}\ll \mathcal{E}^{-1/4},
\end{equation} 
which completes the analysis. 

\subsection{Hickman's boules}
\label{sec:Scalinghickman}
We now briefly discuss the original boules of Hickman, where the vapour is generated from the superheated pool. As anticipated in \S \ref{sec:formulation}, the vapour generation can be described Newton's law of cooling: $j=\mathpzc{h} \Delta T /(L\rho_\mathrm{v})$, where $\mathpzc{h}$ is the heat transfer coefficient and the temperature difference now defined based on the (superheated) pool temperature far away from the drop. In this case, $j$ is approximately constant along the gap. Therefore, proceeding in a similar manner as discussed before in \S\ref{sec:analysis} we now obtain for the vapour thickness in outer region 1: 
\begin{equation}
\frac{1}{2\ \sin{\theta}}\partial_\theta \left[\sin^2{\theta}\ t^3\right]=\frac{6 j \eta_\mathrm{v} R}{\rho g}.
\end{equation}
This is the equivalent of (\ref{eq:ploep}) , now adapted to the Hickman boule. Solving this equation yields 
\begin{equation}
t(\theta)=\left( 2 \frac{1-\cos{\theta}}{\sin^2{\theta}}\right)^{\frac{1}{3}} \left(\frac{6 j\eta_\mathrm{v} R}{\rho g }\right)^{\frac{1}{3}}.
\end{equation}   
From this we deduce the (non-dimensional) vapour layer thickness below the centre of the drop ($\tilde{t}_0=\tilde{t}(0)$) and the vapour thickness as it approaches the neck ($\tilde{t}_\mathrm{exit}=\tilde{t}(\pi/2)$):
\begin{equation}
\tilde{t}(0)= \left(\frac{6 j \eta_\mathrm{v} R}{\rho g \lambda_\mathrm{c}^3}\right)^{\frac{1}{3}}, \quad    \tilde{t}(\pi/2)=1.25992\cdots\left(\frac{6 j \eta_\mathrm{v} R}{\rho g \lambda_\mathrm{c}^3}\right)^{\frac{1}{3}}.
\end{equation} 

An important consequence of this result is that the thickness $t$ scales as $\Delta T ^{1/3}$, which is fundamentally different from the $\Delta T ^{1/4}$ scaling found previously. This new scaling law caries through to the thickness of the neck, according to $t_\mathrm{neck} \sim {t_0}^{6/5}\sim \Delta T ^{2/5}$, see \S \ref{sec:selfsim}.

\section{Finite drop sizes and differing liquids}\label{sec:extend}

Until now we have studied the structure of infinitely large Leidenfrost drops on a liquid bath of equal physical properties. It is of course interesting to extend the results to smaller sized drops and to systems of different liquids. In the limit of small evaporation, $\mathcal{E} \ll 1$, one still finds that the vapour layer is asymptotically thin. Hence, the global shape of the drop is expected to be a ``superhydrophobic" drop on a pool, governed by hydrostatics. Exploiting this idea, we demonstrate that the various scaling laws for the vapour layer are robust, as is confirmed by solving the full problem numerically.

\subsection{Finite drop size}
\label{sec:finite}

\begin{figure}
\centering
\includegraphics[width=0.75\linewidth]{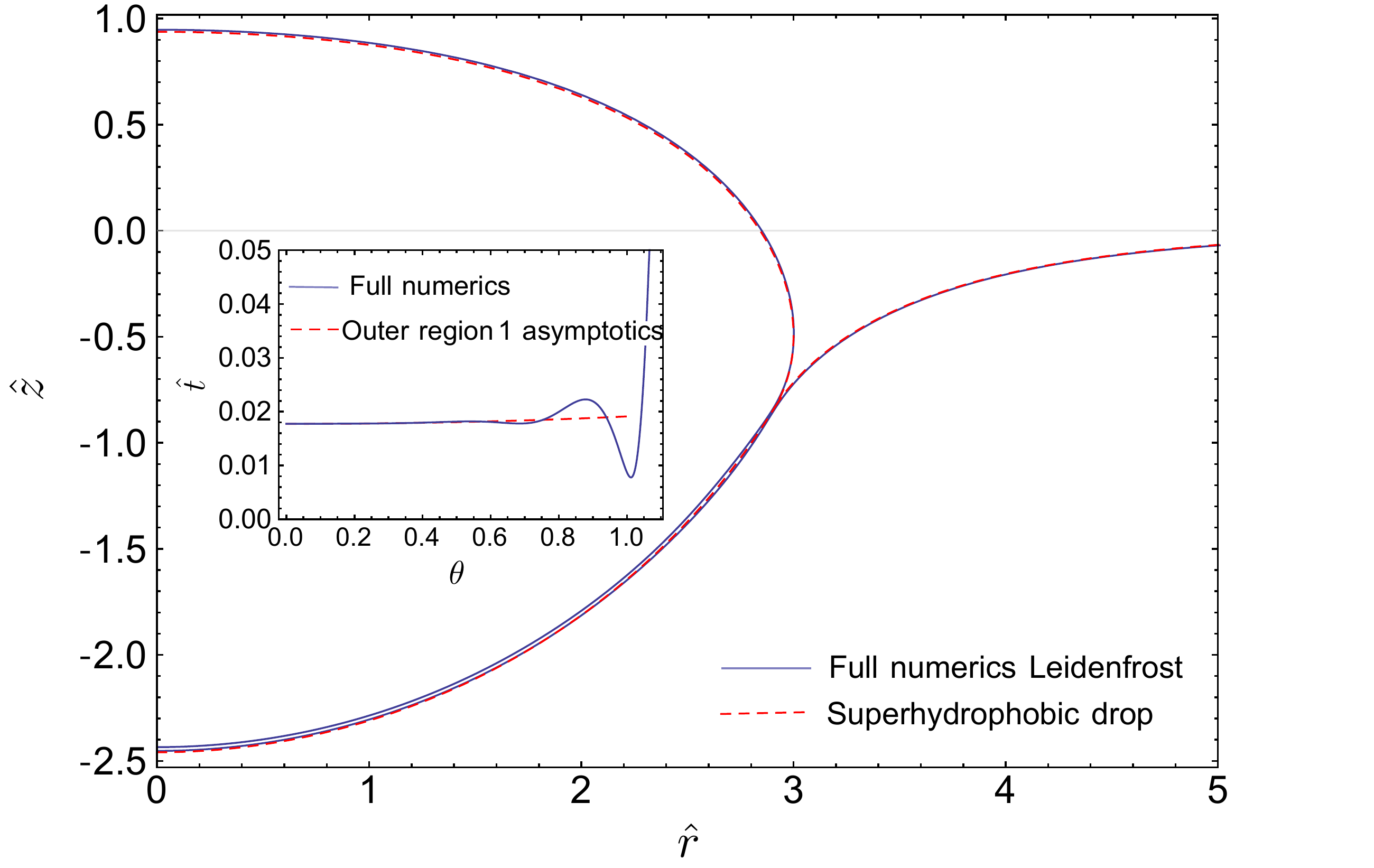}
\caption{Profiles calculated for equal property liquids, $\mathcal{R}=3$ and $\mathcal{E}=8.64\times 10^{-8}$. Both the super-hydrophobic drop calculation and the numerical simulation of the full problem yield a spherical cap solution of curvature $\frac{2}{\mathcal{R}_c}$ for the gap geometry, in agreement with equation (\ref{eq:bla1}). Note that for finite sized drop $\mathcal{R}_c\neq \mathcal{R}$ and the neck is positioned at $\theta_\mathrm{n}=1.01$. }
\label{fig:R3eqprop}
\end{figure}

Let us first focus on finite sized drops, while keeping $\Gamma=\mathcal{P}=1$. The size of the drop can be tuned by $k$ appearing in (\ref{eq:pvdrop}), and a numerical example is presented in blue in \autoref{fig:R3eqprop}. In this particular case the droplet radius $\mathcal R =R/\lambda_\mathrm{c} = 3$ (as seen from above); in general a relation $\mathcal{R}(k)$ can be established numerically (cf.~Appendix~\ref{app:numerics_full}). Comparing the droplet shape to  that of the very large drops in \autoref{fig:sketchzones}, one finds that the immersed part of the drop still resembles a spherical cap, but the position of the neck has clearly shifted, resulting that drop radius $\mathcal R$ is now smaller than the (dimensionless) radius of curvature of the spherical cap, which we define as  $\mathcal{R}_c$. The inset shows details of the vapour layer, which also has a similar structure as compared to large drops at small $\mathcal{E}$. 

These features can be understood in detail. First, we compare the full numerical solution to the reduced (hydrostatic) calculation for the superhydrophobic drop, as described in  \S\ref{sec:asy} (and in more detail in Appendix~\ref{app:superhydrophobic}). The latter is shown as the red dashed curve in \autoref{fig:R3eqprop}, indeed giving an excellent description of the global shape. As a second step, one can use this global shape to predict the gap thickness. Namely, the superhydrophobic drop provides $P_v$ assuming a negligible 
back influence of the vapour film profile on $P_v$; this is valid except for the relatively narrow neck and intermediate regions where the capillary (Laplace) pressure due to vapour film deformation is important. Inserting this pressure profile in the lubrication equation (\ref{eq:general1}), one can obtain the vapour layer profile. The result is shown as the red dashed curve in the inset of \autoref{fig:R3eqprop} and indeed manifests an excellent quantitative agreement, outside the neck region and the oscillatory intermediate region. 

The same asymptotic analysis as for the infinite drop can be applied. However, care must be taken that the oscillation visible in the inset of \autoref{fig:R3eqprop} does not extend all the way to the centre of the drop; otherwise there would be no flat ``outer region" below the drop. Hence, we need the width of the bump, $\hat{\ell}_b \sim \mathcal E^{3/40}$ to remain smaller than $\mathcal R$. We therefore postulate a new hierarchy of lengthscales in the case of finite drop sizes, namely 

\begin{equation}
t_\mathrm{n} \ll t_0 \ll \ell_b\ll R \,,
\label{eq:hierarchy_bis}
\end{equation}
which is the same as $\hat t_\mathrm{n} \ll \hat t_0 \ll \hat{\ell}_b \ll \mathcal{R}$, resulting in 

\begin{equation}
\mathcal{E}^{3/10} \ll \mathcal{E}^{1/4} \ll \mathcal{E}^{3/40} \ll \mathcal{R}.
\end{equation}

Based on these observations we can now revisit the analysis for the vapour layer. For equal material properties and small evaporation numbers, (\ref{eq:bla1}) is still valid so that the immersed part of the drop has a constant curvature. For the numerical example in \autoref{fig:R3eqprop} we find $\lambda_\mathrm{c} \kappa_h=2\times 0.29=2/\mathcal{R}_c$. Also, the lubrication equation (\ref{eq:sphere}) is still valid. However, since the expression for $P_v$ involves $\mathcal{R}_c$ rather than $\mathcal R$, we need to adapt the expression for $t_*=t_0$ accordingly. With this, (\ref{eq:t0}) and the first expression (\ref{eq:t0_texit_dim}) simply become

\begin{equation}\label{eq:t0bis}
t_*=t_0=\lambda_\mathrm{c} \left(\frac{\mathcal{R}_c}{\mathcal R}\right)^{1/4}\mathcal{E}^{1/4},
\end{equation} 
where the ratio $\mathcal{R}_\mathrm{c}/\mathcal{R}$ can be calculated from the corresponding superhydrophobic drops. 
When computing the thickness of the very thin neck, one should take into account two further effects due to the finite drop size. First, the neck is no longer positioned at $\theta=\pi/2$: in the example in \autoref{fig:R3eqprop} we find  $\theta_\mathrm{n}=1.01$, which will lead to a small change in the thickness at the ``exit" of the outer region, $t_{\rm exit}$, in view of the weak variation of $t$ with theta. Second, the matching of the neck to the upper surface of the droplet will be modified, since the droplet's curvature will change with respect to the value for an infinitely large drop. Since all these factors have to be evaluated numerically, we here just give the scaling of the neck thickness

\begin{equation}\label{eq:tnbis}
t_\mathrm{n} \sim \lambda_\mathrm{c} \left(\frac{\mathcal{R}_\mathrm{c}}{\mathcal{R}}\right)^{3/10}\mathcal{E}^{3/10}.
\end{equation}
%\begin{equation}\label{eq:tnbis}
%t_\mathrm{n}=2 \cdot  1.147 \ \left( \frac{\tilde t(\theta_\mathrm{n})}{2} \right)^{6/5} \lambda_\mathrm{c} \left(\frac{\mathcal{R}_c}{\mathcal{R}}\right)^{3/10}\mathcal{E}^{3/10},
%\end{equation}
We also note that the prefactor in this law exhibits some dependence with $\mathcal{R}$ when $\mathcal{R}\approx 1$, and therefore the scaling $\hat t_\mathrm{n}\sim \mathcal{R}^{3/10}$ may not actually hold for $\mathcal{R}\approx 1$. This explains why the apparent scaling identified by \cite{Maquet2016} was rather $\hat t_\mathrm{n}\sim \mathcal{R}^{1/4}$ for the range of radii studied there. 

In summary, we conclude that the structure of the present asymptotic analysis and the resulting scaling laws remain the same for finite sized drops, provided that the undulations near the neck do not penetrate a large fraction of the gap length (i.e.\ provided that the intermediate region stays well shorter than the outer region 1). This being satisfied, in the case of equal liquid properties one can even compute the prefactors, provided that $\mathcal{R}_\mathrm{c}$ and $\theta_\mathrm{n}$ are determined by considering the corresponding superhydrophobic drop. Note that the (dimensional) drop size $R$ appears both in $\mathcal E$ and in the prefactors of (\ref{eq:t0bis}) and (\ref{eq:tnbis}). Hence, as already mentioned, one observes a pure scaling relation in terms of $R$  only in the large drop limit, although this is limited to the case of equal liquid properties.

\subsection{Differing liquids}
\label{sec:differing}
We close the discussion by considering cases when $\Gamma=\gamma_\mathrm{p}/\gamma_\mathrm{d} \neq 1$ and $\mathcal{P}=\rho_\mathrm{p}/\rho_\mathrm{d} \neq 1$, i.e. when the drop and pool consist of a different liquid. This is for example the case in the experiments with ethanol drops on a silicone oil pool by \citet{Maquet2016}, who also provided a direct comparison between theory and experiment. Note that the immersed shape is now no longer expected to be a spherical cap due to the difference in capillary length of the pool and drop given their different densities and/or surface tensions. A consequence of this is that we can no longer find an analytical expression for the vapour gap thickness in closed form as before. However, a detailed quantitative asymptotic analysis can still be performed, with the lowest-order solution being given by the numerical solution for the superhydrophobic drop. With this input we can perform a direct comparison with drop shapes obtained from the full numerical solution. The comparison can be made not only for the global shape of the drop, but also for the vapour film below the drop and in the thin neck region. Namely, the superhydrophobic drop determines $P_v$ which combined with (\ref{eq:general1}) gives a fully quantitative prediction for the film profile. The scaling laws $t_0 \sim \lambda_\mathrm{c} \mathcal{E}^{1/4}$ and $t_\mathrm{n} \sim \lambda_\mathrm{c} \mathcal{E}^{3/10}$ will turn out to be preserved, and the prefactors will be determined numerically from the superhydrophobic drop. We formalize these points below.

We use $\lambda_\mathrm{c}$ as the length scale and $\rho_\mathrm{d} g \lambda_\mathrm{c}=\gamma_\mathrm{d}/\lambda_\mathrm{c}$ as the pressure scale to introduce dimensionless variables, marked by a hat, 
\begin{equation}
\hat{s}=\frac{s}{\lambda_\mathrm{c}}\,,\ \hat{r}=\frac{r}{\lambda_\mathrm{c}}\,,\ \hat{z}=\frac{z}{\lambda_\mathrm{c}}\,,\ \hat{P}_v=\frac{P_v}{\rho_\mathrm{d} g \lambda_\mathrm{c}}\,,\ 
\label{eq:adims}
\end{equation}
similar to earlier used $\hat{t}=t/\lambda_\mathrm{c}$. We remind that $r$ is the radial coordinate, $s$ the curvilinear coordinate along the vapour film while $z$ is the vertical coordinate. 
Equation (\ref{eq:general1}) then rewrites as
\begin{equation}
-\frac{1}{12 \hat{r}} \partial_{\hat{s}}\left(\hat{r} \hat{t}^3 \partial_{\hat{s}} \hat{P}_v\right)=\frac{\tilde{\mathcal{E}}}{\hat{t}} \,.
\label{eq:general1_adim2} 
\end{equation}
In the outer region 1 considered in the framework of our present asymptotic scheme, $\hat{P}_v(\hat{s})$ and $\hat{r}(\hat{s})$ are \textit{a priori} given functions, known from the superhydrophobic drop consideration (cf.~Appendix~\ref{app:superhydrophobic}). Then the solution to (\ref{eq:general1_adim2}), non-singular at the symmetry axis $\hat{s}=0$, can be written as 
\begin{equation}
\hat{t}^4=\frac{16 \tilde{\mathcal{E}} \int_0^{\hat{s}} \hat{r}^{4/3}(-\hat{P}'_v)^{1/3} \,\texttt{d}\hat{s} }{(-\hat{r}\,\hat{P}'_v)^{4/3}} \,,
\label{eq:profile_outer1}
\end{equation}
the prime denoting a derivative with respect to $\hat{s}$. 
The thickness at the symmetry axis, $\hat{t}_0=\hat{t}(0)$, and at the exit, $\hat{t}_\text{exit}=\hat{t}(\hat{s}_\text{CL})$, can therefrom be inferred as
\begin{equation}
\hat{t}_0=\left(-\frac{6\,\tilde{\mathcal{E}}}{\hat{P}''_v(0)}\right)^{1/4} \,, \quad \hat{t}_\text{exit}=\frac{2\,\tilde{\mathcal{E}}^{1/4}\left(\int_0^{\hat{s}_\text{CL}} \hat{r}^{4/3} (-\hat{P}'_v)^{1/3} \,\texttt{d}\hat{s}
\right)^{1/4}}{[-\hat{r}(\hat{s}_\text{CL}) \hat{P}'_v(\hat{s}_\text{CL})]^{1/3}} \,,
\label{eq:t_extremities}
\end{equation}
where $\hat{s}_\text{CL}$ is the value of the arc length at the contact line of the superhydrophobic drop (known \textit{a priori}, cf.~Appendix~\ref{app:superhydrophobic}). We note that $\hat{P}'_v<0$ and $\hat{P}''_v(0)<0$. We also note that the results of \S\ref{sec:analysis} with a hemispherical shape and of \S\ref{sec:finite} with a spherical cap shape are recovered from here with $\texttt{d}\hat{s}=\mathcal{R}\,\texttt{d}\theta$, $\hat{P}_v'=-\sin\theta$, $\hat{r}=\mathcal{R} \sin\theta$ and $\hat{s}_\text{CL}=\mathcal{R}\,\theta_\text{CL}$. Equations (\ref{eq:profile_outer1}) and (\ref{eq:t_extremities}) confirm once again that the scaling in terms of $\mathcal{E}$ established in \S\ref{sec:analysis} for the vapour gap thickness in this outer region~1, namely $\hat{t}=O(\tilde{\mathcal{E}}^{1/4})$ including $\hat{t}_0=O(\tilde{\mathcal{E}}^{1/4})$ and $\hat{t}_\text{exit}=O(\tilde{\mathcal{E}}^{1/4})$, is indeed robust.  

Turning to the neck region, we introduce a local Cartesian coordinate $x$ (and its dimensionless version $\hat{x}=x/\lambda_\mathrm{c}$) parallel to the slope of the superhydrophobic drop at its contact line and pointing away from the vapour film. In this local Cartesian system, the drop and pool surfaces are described by $\hat{d}(\hat{x})$ and $\hat{p}(\hat{x})$, respectively, while $\hat{t}=\hat{d}-\hat{p}$ and $\texttt{d}\hat{x}\approx \texttt{d}\hat{s}$. Owing to the expected small size of the neck region, the leading-order contributions into $\partial_{\hat{x}}\hat{P}_v$ will be due to the capillary (Laplace) pressure associated with the first curvature of the drop and pool surfaces, i.e.\ $\hat{P}'_v\approx -\hat{d}'''$ and $\hat{P}'_v\approx \Gamma\hat{p}'''$. The latter two expressions must be equal, hence $\hat{d}'''\approx -\Gamma\hat{p}'''$. Using this fact, as well as $\hat{t}=\hat{d}-\hat{p}$, we express $\hat{P}'_v\approx -\frac{\Gamma}{1+\Gamma}\hat{t}'''$. Using this in (\ref{eq:general1_adim2}) and recalling, on the one hand, that it is planar geometry that holds to leading order in the neck region, and on the other hand, that the local evaporation flux $\tilde{E}/\hat{t}$ is negligible relative to the flux from the remainder of the vapour gap passing through the neck region (cf.~\S\ref{sec:inner}), we obtain 
\begin{equation}
\partial_{\hat{x}}
\left(\hat{t}^3 \partial_{\hat{x}\hat{x}\hat{x}}\hat{t}\right) = 0
\label{eq:neck1}
\end{equation}
to leading order in the neck region. 

As in \S\ref{sec:inner}, we shall look for solutions of this equation subject to boundary conditions $\hat{t}''(-\infty)=0$ whilst $\hat{t}''(+\infty)=\hat{\kappa}_{1h,\text{CL}}-\hat{\kappa}_{1e,\text{CL}}$, where $\hat{\kappa}_{1h,\text{CL}}$ and $\hat{\kappa}_{1e,\text{CL}}$ are the \textit{first} curvatures of the upper drop and pool surfaces at the contact line of the superhydrophobic drop, known from the corresponding superhydrophobic drop solution (cf.~Appendix~\ref{app:superhydrophobic}). Recall that in the particular case of large drops with the same liquid properties (cf.~\S\ref{sec:inner}), we have $\hat{\kappa}_{1h,\text{CL}}=\sqrt{2}$ and $\hat{\kappa}_{1e,\text{CL}}=-\sqrt{2}$ resulting from the puddle solution, and giving rise to $\hat{t}''(+\infty)=2\sqrt{2}$ in present terms. 

However, these boundary conditions are still insufficient for (\ref{eq:neck1}). We must also account for the flux coming from the interior of the vapour gap. To this purpose, we integrate (\ref{eq:neck1}) on $\hat{x}$ from $-\infty$ to a finite value of $\hat{x}$. When evaluating the resulting terms at $\hat{x}=-\infty$, we assume that the flux coming through the neck is entirely determined by the outer region 1, with a possible contribution from the intermediate region being negligible to leading order. We thereby arrive at 
\begin{equation}
\hat{t}^3 \hat{t}'''=\frac{1+\Gamma}{\Gamma} \hat{t}_\text{exit}^3 [-\hat{P}'_v(\hat{s}_\text{CL})]\equiv 16 c^3 \,, 
\label{eq:neck3}
\end{equation}
where $\hat{t}_\text{exit}$ is given by (\ref{eq:t_extremities}), while $\hat{P}'_v(\hat{s}_\text{CL})$ is a value known from the superhydrophobic drop consideration. The definition of $c$ introduced here for the sake of brevity can be seen to coincide in the limit of a hemispherical drop with the one used in \S\ref{sec:inner}. A rescaling 
$$\hat{t}=\left(16 c^3\right)^{2/5} \text{factor}_{\kappa}^{-3/5}\, T\,,\qquad  
\hat{x}=\left(16 c^3\right)^{1/5} \text{factor}_{\kappa}^{-4/5}\, \zeta\,,$$
where $\text{factor}_{\kappa}\equiv (\hat{\kappa}_{1h,\text{CL}}-\hat{\kappa}_{1e,\text{CL}})/\sqrt{2}$, 
reduces equation (\ref{eq:neck3}) with the earlier mentioned boundary conditions to the problem $T^3 T'''=1$ with $T''(-\infty)=0$ and $T''(+\infty)=\sqrt{2}$ already considered in \S\ref{sec:inner}. In particular, for the minimum neck thickness, the value $T_\mathrm{n}=1.147$ was obtained, which in present terms corresponds to 
\begin{equation}
\hat{t}_\mathrm{n}=1.147\,\left(\frac{1+\Gamma}{\Gamma} \hat{t}_\text{exit}^3 [-\hat{P}'_v(\hat{s}_\text{CL})]\right)^{2/5} \left(\frac{\hat{\kappa}_{1h,\text{CL}}-\hat{\kappa}_{1e,\text{CL}}}{\sqrt{2}} \right)^{-3/5}\,.
\label{eq:neck_min}
\end{equation} 
As $\hat{t}_\text{exit}=O(\tilde{\mathcal{E}}^{1/4})$ (cf.~above) with the other quantities in (\ref{eq:neck_min}) being just $O(1)$, we see that $\hat{t}_\mathrm{n}=O(\tilde{\mathcal{E}}^{3/10})$, which confirms the robustness of the earlier established scaling law for the neck thickness. 

With the $\tilde{\mathcal{E}}$ scaling (power) laws themselves confirmed, 
we recall that the associated prefactors coming from the asymptotic results such as (\ref{eq:profile_outer1}), (\ref{eq:t_extremities}) and (\ref{eq:neck_min}) depend exclusively on characteristics of the corresponding superhydrophobic drop. The latter are computed numerically as described in Appendix~\ref{app:superhydrophobic} and subsequently used in (\ref{eq:profile_outer1}), (\ref{eq:t_extremities}) and (\ref{eq:neck_min}) to complete the present asymptotic consideration. The results are illustrated below together with their comparison with the full numerics (the latter realised as described in Appendix~\ref{app:numerics_full}). 

%\begin{figure}
%\resizebox{1\columnwidth}{!}{% 
%\includegraphics{globalR3PGmaquetEsmall.pdf}\qquad\qquad\includegraphics{globalR3PGmaquetElarge.pdf}}\\ \vspace{0.03in} \\
%\resizebox{1\textwidth}{!}{%
%\includegraphics{filmR3PGmaquet.pdf}\qquad\qquad
%\includegraphics{scalingsR3PGmaquet.pdf}}
%\caption{Global drop shapes (top panels) and vapour gap thicknesses (bottom left panel: distributions along the arc length of the pool surface; bottom right panel: values in the center and at the neck as functions of $\tilde{\mathcal{E}}$). Results for $\mathcal{R}=3$ and the material properties resembling those for ethanol drop on silicone oil pool \citep{Maquet2016}. In particular, $\mathcal{P}=1.244$ and $\Gamma=1.156$.  The top panels are for 
%$\tilde{\mathcal{E}}=4.8\times 10^{-9}$ ($\Delta T=1\,\text{K}$) and $\tilde{\mathcal{E}}=2.3\times 10^{-7}$ ($\Delta T=40\,\text{K}$), respectively. And so are the lower and upper curves in the left bottom panel. The arc length $\hat{s}$ is along the pool surface in the full numerics, while along the superhydrophobic drop--pool interface in the outer region 1 asymptotics. }
%\label{fig:R3PGmaquet}
%\end{figure}

\begin{figure}
\includegraphics[width=\linewidth]{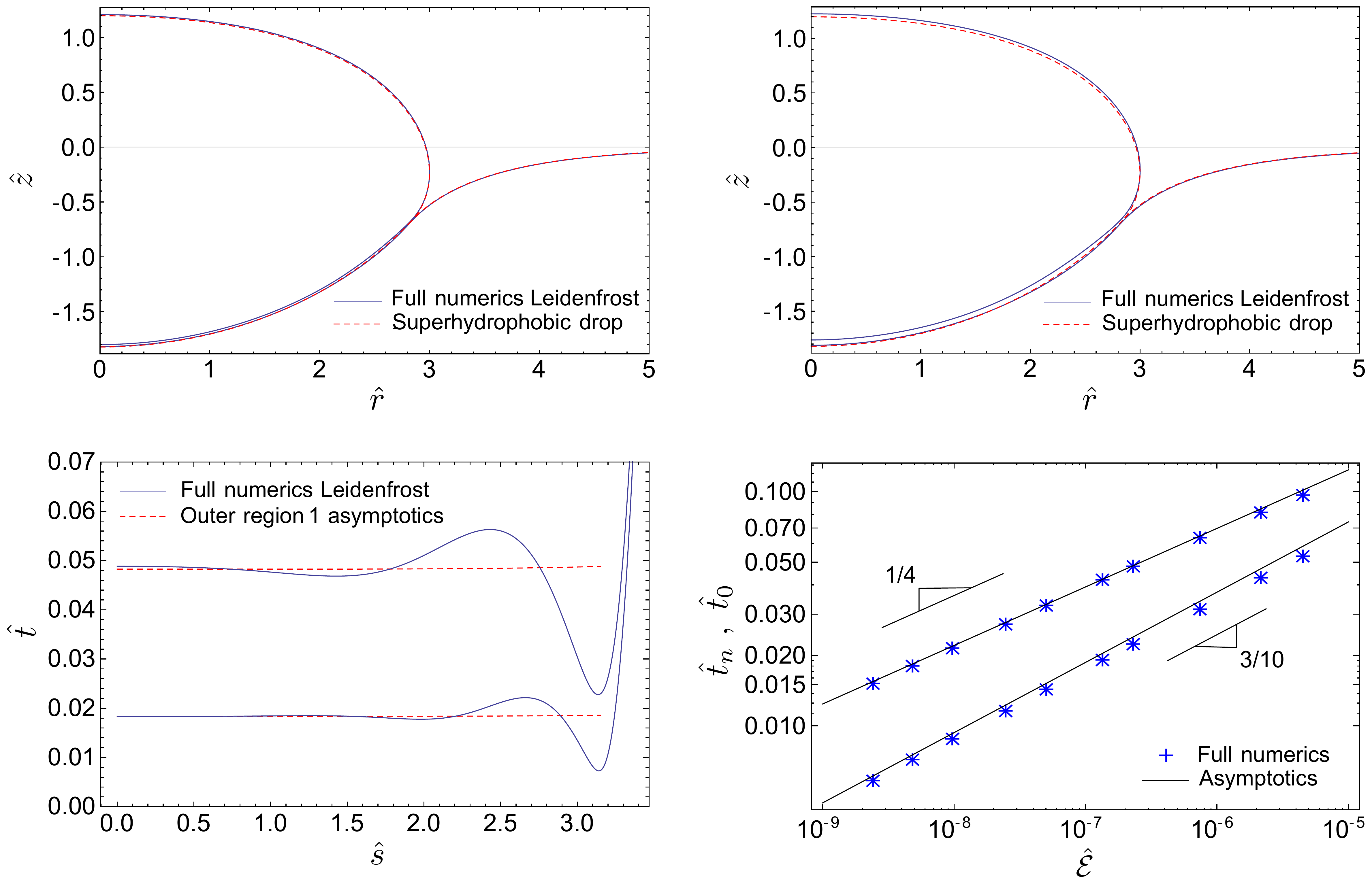}
\caption{Global drop shapes (top panels) and vapour gap thicknesses (bottom left panel: distributions along the arc length of the pool surface; bottom right panel: values in the center and at the neck as functions of $\tilde{\mathcal{E}}$). Results for $\mathcal{R}=3$ and the material properties resembling those for ethanol drop on silicone oil pool \citep{Maquet2016}. In particular, $\mathcal{P}=1.244$ and $\Gamma=1.156$.  The top panels are for 
$\tilde{\mathcal{E}}=4.8\times 10^{-9}$ ($\Delta T=1\,\text{K}$) and $\tilde{\mathcal{E}}=2.3\times 10^{-7}$ ($\Delta T=40\,\text{K}$), respectively. And so are the lower and upper curves in the left bottom panel. The arc length $\hat{s}$ is along the pool surface in the full numerics, while along the superhydrophobic drop--pool interface in the outer region 1 asymptotics. }
\label{fig:R3PGmaquet}
\end{figure}

%In the figures below we use $\lambda_\mathrm{c}$ as the length scale and introduce dimensionless variables, marked by a hat, 
%\begin{equation}
%\hat{r}=\frac{r}{\lambda_\mathrm{c}}\,,\ \hat{z}=\frac{z}{\lambda_\mathrm{c}}\,, \hat{s}=\frac{s}{\lambda_\mathrm{c}}\,\ 
%\label{eq:adims}
%\end{equation}
%similar to earlier used $\hat{t}=t/\lambda_\mathrm{c}$. We remind that $r$ is the radial coordinate, $s$ the curvilinear coordinate along the vapour film while $z$ is the vertical coordinate. 

Figure~\ref{fig:R3PGmaquet} shows results for a drop of $\mathcal{R}=3$ with non-equal material properties inspired from \citet{Maquet2016}. We see that the Leidenfrost drop and pool shapes are still close to those of the superhydrophobic solutions, as expected. The structure of the vapour gap is still the same as noted earlier in the case of equal material properties. The asymptotic results for the outer region 1 capture well the film thickness distribution in the central part, although at larger $\tilde{\mathcal{E}}$ the waviness from the intermediate region penetrates closer to the symmetry axis. For the thickness values in the center and at the neck, we once again obtain a good agreement bewteen the asymptotic and the full numerical approaches, although the agreement slightly deteriorates at larger $\tilde{\mathcal{E}}$ (especially for $\hat{t}_\mathrm{n}$). Importantly, the asymptotic scalings $\hat{t}_0\sim \tilde{\mathcal{E}}^{1/4}$ and $\hat{t}_\mathrm{n}\sim \tilde{\mathcal{E}}^{3/10}$ are seen to still be well reproduced by the full numerics. We note that the straight lines in the right bottom panel of figure~\ref{fig:R3PGmaquet} are the asymptotic predictions with no fitting involved: the prefactors were determined directly from the superhydrophobic drop analysis. 

%\begin{figure}
%\resizebox{1\textwidth}{!}{% 
%\includegraphics{globalRseriesPGmaquetElarge.pdf}}\\ 
%\vspace{0.03in} \\
%\resizebox{0.98\textwidth}{!}{%
%\includegraphics{filmRseriesPGmaquetElarge.pdf}}
%\caption{Extension of the part of the results of figure~\ref{fig:R3PGmaquet} pertaining to $\tilde{\mathcal{E}}=2.3\times 10^{-7}$ (still with $\mathcal{P}=1.244$, $\Gamma=1.156$) to a number of drop radii, $\mathcal{R}=1.5,\,2,\,3,\,10$. In the bottom panel, the correspondence between each curve and the $\mathcal{R}$ value can reasonably be guessed from the right ending of the former.}
%\label{fig:RseriesPGmaquet}
%\end{figure}

\begin{figure}
\centering
\includegraphics[width=0.75\linewidth]{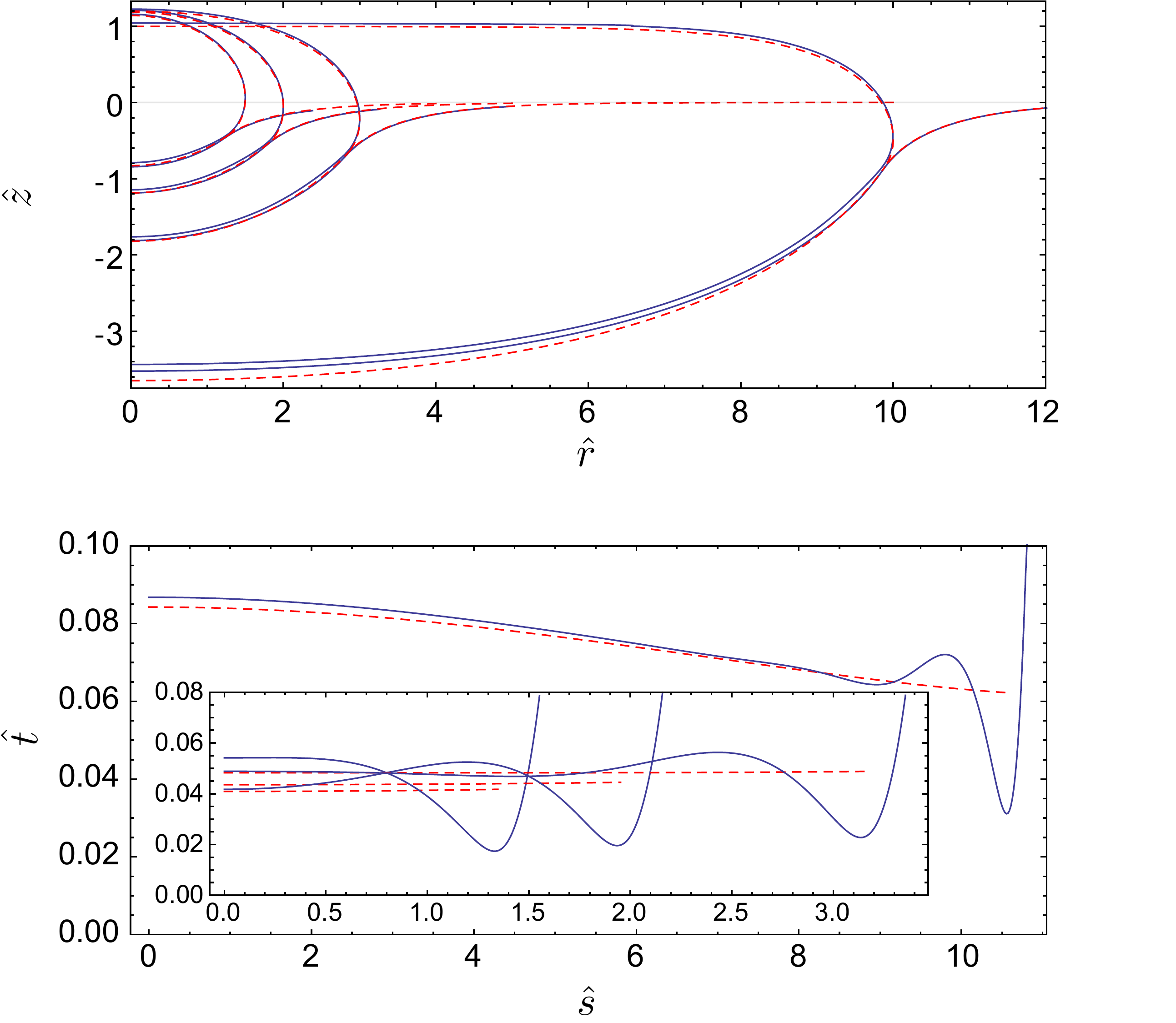}
\caption{Extension of the part of the results of figure~\ref{fig:R3PGmaquet} pertaining to $\tilde{\mathcal{E}}=2.3\times 10^{-7}$ (still with $\mathcal{P}=1.244$, $\Gamma=1.156$) to a number of drop radii, $\mathcal{R}=1.5,\,2,\,3,\,10$. In the bottom panel, the correspondence between each curve and the $\mathcal{R}$ value can reasonably be guessed from the right ending of the former.}
\label{fig:RseriesPGmaquet}
\end{figure}

\begin{figure}
\includegraphics[width=\linewidth]{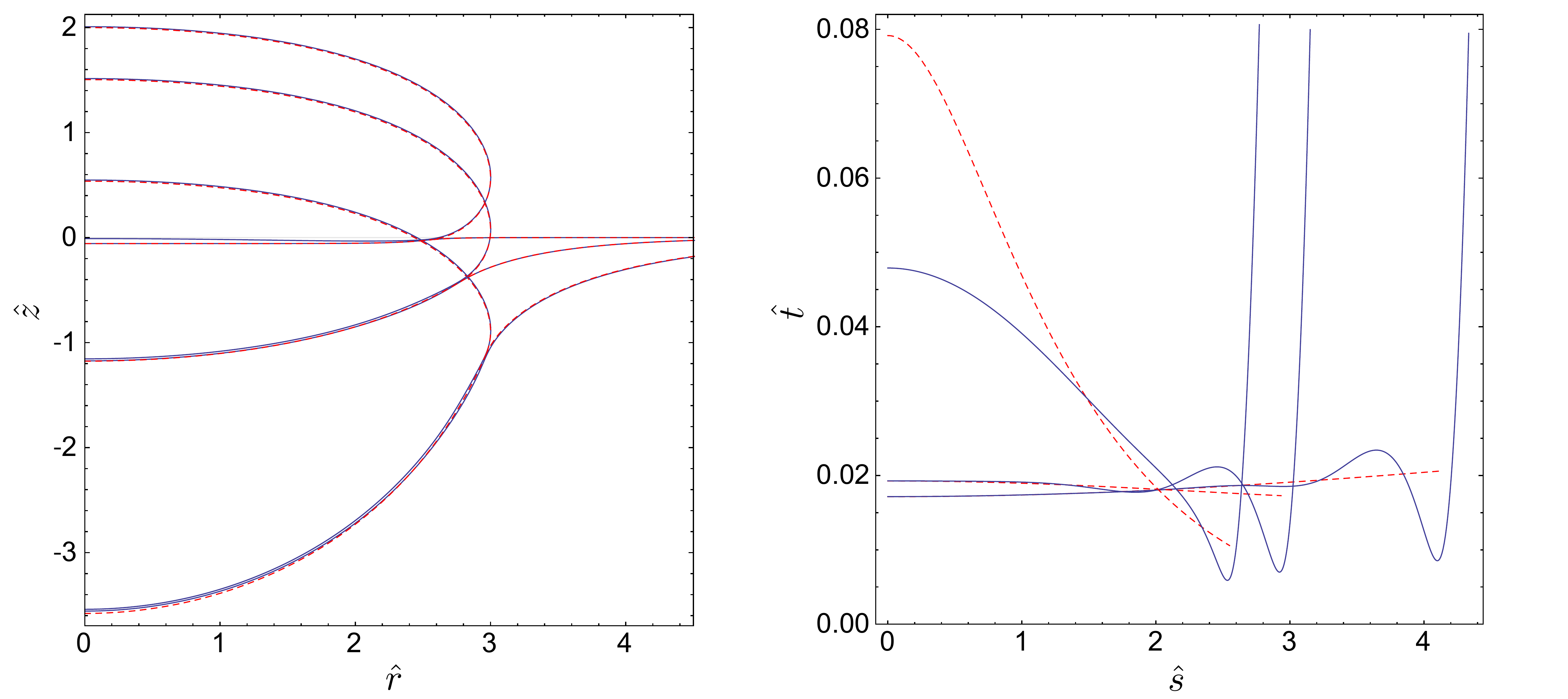}
\caption{Extension the results of figure~\ref{fig:R3eqprop} to a number of pool-to-drop density ratios $\mathcal{P}=0.8,\,2,\,40$ (otherwise still for $\mathcal{R}=3$, $\Gamma=1$ and $\tilde{\mathcal{E}}=4.8\times 10^{-9}$, which corresponds to $\mathcal{E}=8.64\times 10^{-8}$). As $\mathcal{P}$ is increased, the drop mounts in the left panel, while the curves in the right panel get shorter from the right (the overall arc length of the vapour gap decreases).  }
\label{fig:R3PseriesG1}
\end{figure}

Some of the cases shown in figures~\ref{fig:R3eqprop} and~\ref{fig:R3PGmaquet} are taken up for a further parametric study in figures~\ref{fig:RseriesPGmaquet} and~\ref{fig:R3PseriesG1}, partly aimed at testing the limits of the applicability of the present asymptotic scheme. In particular, the results of figure~\ref{fig:R3PGmaquet} corresponding to the larger evaporation number value ($\tilde{\mathcal{E}}=2.3\times 10^{-7}$) are extended in figure~\ref{fig:RseriesPGmaquet} to some other drop radius values. While the Leidenfrost drops are seen to still be close to the superhydrophobic shapes in all cases shown, the present asymptotic results cease to be valid for smaller drops as far as the outer region 1 and $\hat{t}_0$ are concerned. This is especially true for the smallest drop displayed, with $\mathcal{R}=1.5$. It can clearly be observed from the bottom panel of figure~\ref{fig:RseriesPGmaquet} that such a change in the vapour gap morphology is related first of all with the violation of the part $\hat{l}_\mathrm{n}\ll\mathcal{R}$ of the presumed hierarchy of length scales (\ref{eq:hierarchy_bis}). This violation occurs for larger $\tilde{\mathcal{E}}$ as $\mathcal{R}$ is decreased, when the wavelength of the intermediate-region undulations becomes comparable with the size of the drop, as already discussed at the end of \S\ref{sec:finite}. 
At smaller $\tilde{\mathcal{E}}$ though (e.g. at $\tilde{\mathcal{E}}=4.8\times 10^{-9}$ used earlier in figure~\ref{fig:R3PGmaquet}), the present asymptotic scheme is still found to work rather satisfactorily 
in the central part of the drop even for $\mathcal{R}$ as low as $\mathcal{R}=1.5$ (the result not shown). 

A no less remarkable result of figure~\ref{fig:RseriesPGmaquet} is some deterioration of the agreement between asymptotics and full numerics observed for large $\mathcal{R}$ (viz.~$\mathcal{R}=10$) as compared to $\mathcal{R}=3$ as far as the gap thickness in the center (i.e.~$\hat{t}_0$) is concerned. There is a good reason for that. Indeed, it is clear that, quite unlike a superhydrophobic drop over a pool with equal liquid properties, a drop of a differing liquid will tend to adopt a puddle-like shape as $\mathcal{R}$ is increased (here limiting consideration to the case $\mathcal{P}>1$, of a lighter drop; a heavier one will merely sink for sufficiently large $\mathcal{R}$). For such a puddle, which the drop with $\mathcal{R}=10$ of figure~\ref{fig:RseriesPGmaquet} already much resembles, it is not only the upper surface that flattens, but also the immersed one. The present asymptotic scheme breaks down in the presence of a flattened part of the immersed surface, for which the $\hat{P}_v'$ and $\hat{P}_v''$ values to be used in (\ref{eq:profile_outer1}) and (\ref{eq:t_extremities}) vanish and the driving force of the flow $\hat{P}_v'$ is no longer accurately estimated by the superhydrophobic drop. The length scale at which such flattening occurs is the (dimensionless) capillary length of the immersed surface $\sqrt{(1+\Gamma)/(\mathcal{P}-1)}$. Hence the present asymptotic scheme is expected to work in an optimal way when $\mathcal{R}\lesssim \sqrt{(1+\Gamma)/(\mathcal{P}-1)}$, i.e.\ for sufficiently small drops. The drop with $\mathcal{R}=10$ of figure~\ref{fig:RseriesPGmaquet} is already relatively large in this regard, hence the mentioned agreement deterioration. Note that this concerns a relatively larger value $\tilde{\mathcal{E}}=2.3\times 10^{-7}$. For a smaller value $\tilde{\mathcal{E}}=4.8\times 10^{-9}$, the agreement still proves to be excellent even for $\mathcal{R}=10$ (the result not shown). Note also that for a system with $\mathcal{P}=1$, the capillary length of the immersed surface goes to infinity (the surface itself assuming the form of a spherical cap), and hence such a limitation on the maximum drop radius is lifted.

%\begin{figure}
%\resizebox{1\textwidth}{!}{% 
%\includegraphics{globalR3PseriesG1Esmall.pdf}\qquad 
%\includegraphics{filmR3PseriesG1Esmall.pdf}}
%\caption{Extension the results of figure~\ref{fig:R3eqprop} to a number of pool-to-drop density ratios $\mathcal{P}=0.8,\,2,\,40$ (otherwise still for $\mathcal{R}=3$, $\Gamma=1$ and $\tilde{\mathcal{E}}=4.8\times 10^{-9}$, which corresponds to $\mathcal{E}=8.64\times 10^{-8}$). As $\mathcal{P}$ is increased, the drop mounts in the left panel, while the curves in the right panel get shorter from the right (the overall arc length of the vapour gap decreases).  }
%\label{fig:R3PseriesG1}
%\end{figure}

In figure~\ref{fig:R3PseriesG1}, we take up the case of figure~\ref{fig:R3eqprop} (with $\mathcal{R}=3$, $\Gamma=1$ and $\tilde{\mathcal{E}}=4.8\times 10^{-9}$) and explore how the phenomenon is affected by the variation of the density ratio $\mathcal{P}$. We see that, as long as the pool surface deformation remains appreciable (e.g.~for $\mathcal{P}=0.8,\,2$), the expected morphological features remain in place. In particular, this is still the case at $\mathcal{P}=0.8$, for nearly the maximum possible immersed-surface deformation: we are then on the verge of the disappearance of a floating drop configuration for $\mathcal{P}$ slightly below $\mathcal{P}=0.8$, when a drop with $\mathcal{R}=3$ and $\Gamma=1$ becomes too heavy to remain on the pool surface. 
Remarkably, the fact that such disappearance happens practically simultaneously (in terms of $\mathcal{P}$) for the Leidenfrost and superhydrophobic drops can serve as yet another confirmation of a close relationship between these two configurations. On the other hand, while perhaps not physically relevant, the interest of the case $\mathcal{P}=40$ is that the pool surface becomes essentially flat: hence we should thereby approach the case of a Leidenfrost drop above a (flat) solid. Indeed, the vapour film profile for $\mathcal{P}=40$ is seen (cf.~the right panel of figure~\ref{fig:R3PseriesG1}) to closer resemble the one typical for a flat solid substrate~\citep{Snoeijer2008,Sobac2014}, with the present asymptotic scheme yielding an appreciable error with respect to the full numerics. The reason for such an error can be traced back to the fact that, under the conditions when the vapour gap thickness becomes quite comparable with (or even larger than) the deviation of the immersed surface from horizontality, the pressure distribution $\hat{P}_v(\hat{s})$ can no longer be taken as $\hat{t}$-independent as assumed from the superhydrophobic drop considerations. There must rather be a feedback between $\hat{P}_v(\hat{s})$ and $\hat{t}(\hat{s})$ as the one in the case of a flat solid substrate~\citep{Snoeijer2008,Sobac2014}. 
Note also that a similar observation can actually be attributed to the flattened parts (if any) of the immersed surface even if the latter is not flattened overall 
(cf.\ the above discussion of the drop with $\mathcal{R}=10$ of figure~\ref{fig:RseriesPGmaquet}). At last, it is needless to say that an increase of the pool-to-drop surface tension ratio $\Gamma$ 
will lead to similar effects of pool surface flattening as an increase of $\mathcal{P}$ does. 

Thus, for differing liquids too, 
the scalings $\hat{t}_0\sim \tilde{\mathcal{E}}^{1/4}$ and $\hat{t}_\mathrm{n}\sim \tilde{\mathcal{E}}^{3/10}$ are still robust and work in a reasonable interval of small $\tilde{\mathcal{E}}$ values for sufficiently large drops ($\mathcal{R}\gg\hat{l}_\mathrm{n}$) 
in the general case of well curved pool surfaces. However, the particular cases when the immersed superhydrophobic drop surface undergoes a strong overall (for too large $\mathcal{P}$ and/or $\Gamma$) or partial (for too large $\mathcal{R}$) flattening must realistically be excluded from the applicability domain of the present asymptotic scheme. Indeed, such applicability would then just be expected for extremely small values of $\tilde{\mathcal{E}}$, of no practical interest. In contrast, the pure scaling in terms of $\mathcal{R}$ underscored in \S\ref{sec:finite} 
makes actually no sense as such here (for differing liquids)  
due to a morphological difference between the superhydrophobic drop shapes in the limit of large $\mathcal{R}$: a hemisphere for equal liquid properties, and a puddle for differing ones (assuming a lighter drop, $\mathcal{P}>1$) with further adverse consequences due to the mentioned partial flattening of the immersed surface.

\section{Conclusion}\label{sec:Discussion}

Leidenfrost drops on a superheated liquid pool were studied in the limit of small evaporation numbers $\tilde{\mathcal{E}}$, the latter proportional to the superheat $\Delta T$ and determined by both thermal and hydrodynamic properties of the system. The pool surface being deformed under the drop, the vapour gap was found to be of quite a different morphology as compared to that of Leidenfrost drops deposited on a superheated flat plate. 
The reason is that, for a curved substrate, there exists an \textit{a priori} given driving pressure gradient in the vapour layer, determined by the associated superhydrophobic drop configuration and independent to leading order from the gap thickness distribution. 
In contrast, for a flat substrate, the pressure gradient is fully determined by the variation of the gap thickness. 

As shown in detail in \autoref{fig:sketchzones}, three main asymptotic regions are identified to describe the thin gap of vapour between the droplet and the liquid pool. This is unlike the case for the Leidenfrost drop on a flat substrate, for which only two regimes appear.
  
First, an outer region is identified, which is asymptotically the longest one of the three. Its longitudinal extent is comparable with the size of the drop. The vapour gap thickness in this region, scaling as $O(\tilde{\mathcal{E}}^{1/4})$, only marginally varies relative to its value at the symmetry axis. This is quite different from the outer region in the vapour gap over a flat substrate, which is there in the form of a vapour pocket, wide in the center and narrowing towards the edges \citep{Snoeijer2009,Sobac2014}, with a thickness scaling as $O(\tilde{\mathcal{E}}^{1/6})$ \citep{Sobac2014}. It is such difference in morphology that is behind the expected suppression of the chimney instability for large Leidenfrost drops over a liquid pool, unlike their counterparts over a flat substrate \citep{Snoeijer2009}. 
An analytical solution for the thickness profile in the outer region was found in the case of large drops of the same liquid properties as the pool.  In the case of smaller drops and/or differing properties, such a solution is expressed through numerically determined characteristics of the associated superhydrophobic drop. 

Second, we identified an inner (neck) region at the exit from the vapour gap, the only one that bears a great resemblance with the corresponding region for Leidenfrost drops on a flat substrate \citep{Snoeijer2009,Sobac2014}. The neck thickness is here found to scale as $O(\tilde{\mathcal{E}}^{3/10})$, which is close to $O(\tilde{\mathcal{E}}^{1/3})$ obtained for a flat substrate \citep{Sobac2014}. The longitudinal extent of the neck scales as $O(\tilde{\mathcal{E}}^{3/20})$ (cf.\ $O(\tilde{\mathcal{E}}^{1/6})$ for a flat substrate). A self-similar structure was found for the neck profile, which turns out the be governed by the same universal solution as for the Leidenfrost drop on a rigid substrate. The scaling laws and the self-similar shape showed excellent agreement with the full numerical solution.

Third, the peculiarities of the outer region morphology over a curved pool surface yet require the existence of an intermediate region to join the inner (neck) and outer ones. In contrast, no such intermediate region is present in the vapour gap over a flat substrate \citep{Snoeijer2009,Sobac2014}, where the outer and inner regions directly match one another. The vapour gap profile in this intermediate region turns out to be in the form of stationary capillary waves (undulations) springing from the neck and decaying towards the outer region. The largest bump, next to the neck, scales as $O(\tilde{\mathcal{E}}^{9/40})$ in thickness, and $O(\tilde{\mathcal{E}}^{3/40})$ in longitudinal extent. It is partly an excessive penetration of the undulations into the outer region that, for small but finite $\tilde{\mathcal{E}}$, limits the applicability of our asymptotic scheme as far as sufficiently small drops are concerned. 

On the one hand, the robustness of our analysis was tested in the realm of finite-sized (but not too small) drops and differing liquids with a largely positive outcome, provided that the parameters are such that the pool surface remains essentially curved.  
 On the other hand, the case of a large drop with the same liquid properties as the pool stands out  as the ideal baseline for our asymptotic scheme and for which a large part of the analysis was carried out. Importantly, for these large Leidenfrost drops, pure scaling relations can be established in terms of the drop radius $R$ too, and not just in terms of $\tilde{\mathcal{E}}$ (a size-independent quantity).  
To recover their due form, it suffices to replace $\tilde{\mathcal{E}}$ in the earlier mentioned scaling relations with a modified (size-dependent) evaporation number defined in the present paper as $\mathcal{E}=6\, \tilde{\mathcal{E}} R/\lambda_\mathrm{c}$. 

It is worth mentioning that the ``boules'' described by \citet{Hickman64}, reproduced in \autoref{fig:HickBoule}, exhibit slightly different scaling laws. This is due to the different way the vapour is generated, which for the Hickman boules comes from the superheated pool. This is to be contrasted with ``usual '' Leidenfrost case, for which the drop is evaporating. The main feature is that $t\sim \Delta T ^{1/3}$ as opposed to $\Delta T ^{1/4}$.

Finally, we found that the configuration of a flat substrate can here be recovered in the case of strongly differing liquid properties and hence a gradual change can in principle be realised between the two systems. However, as far as the details of a gradual transition between the asymptotic paradigms of flat and curved substrates are concerned, they remain a subject of future studies.

\subsection*{Acknowledgments}
Part of this research has been funded by ESA-BELSPO via the Prodex Heat Transfer project, and by BELSPO via the IAP 7/38 MicroMAST network. BS and PC gratefully acknowledge financial support from the Fonds de la Recherche Scientifique - F.N.R.S. This work was partially supported by an ERC advanced grant.

\appendix
\section{Numerical approach to the full problem}
\label{app:numerics_full}

Here we provide details of the numerical approach used to solve the full problem formulated in \S\ref{sec:model}. In essence, the approach is similar to the one used by \citet{Sobac2014,Maquet2016}, the modifications being largely of a geometric nature aimed at dealing with strongly curved substrates. Hereafter, the subscripts `$h$' and `$e$' are used for the geometric quantities belonging to the drop and pool surfaces, respectively. We shall work in dimensionless terms by using $\lambda_\mathrm{c}$ as the length scale and $\rho_\mathrm{d} g \lambda_\mathrm{c}=\gamma_\mathrm{d}/\lambda_\mathrm{c}$ as the pressure scale. The thereby obtained dimensionless variables are marked by hat, as e.g.\ in (\ref{eq:adims}). 
%$$\hat{h}=\frac{h}{\lambda_\mathrm{c}}\,\ \hat{e}=\frac{e}{\lambda_\mathrm{c}}\,,\ \hat{t}=\frac{t}{\lambda_\mathrm{c}}\,\ \hat{r}=\frac{r}{\lambda_\mathrm{c}}\,,\ \hat{s}=\frac{s}{\lambda_\mathrm{c}}\,,\ \hat{\kappa}=\kappa\lambda_\mathrm{c}\,,\ \hat{P}_v=\frac{P_v}{\rho_\mathrm{d} g \lambda_\mathrm{c}}\,,\ \hat{k}=\frac{k}{\rho_\mathrm{d} g \lambda_\mathrm{c}}\,.$$
The only exception to this hat rule is the drop radius $R$, for the dimensionless version of which we already have a notation $\mathcal{R}$, cf.~(\ref{numbers}).  
Nonetheless, expecting no confusion in the reader, the hats will be omitted in the remainder of this appendix for the sake of brevity, whereas they are still retained for the same quantities in the main body of the text. 

\subsection{General equations}

Then equations (\ref{eq:pvpool}), (\ref{eq:pvdrop}), (\ref{eq:general1}) and (\ref{eq:staticshapes}) adopt the following dimensionless forms:  
\begin{eqnarray}
P_v&=& k-h-\kappa_h \qquad \text{(everywhere)}\,, \label{eq:general2_adim} \\
P_v&=&-\mathcal{P}\, e+\Gamma \kappa_e \qquad \text{(everywhere)}\,, \label{eq:general3_adim} \\
P_v&=&0 \qquad\quad \text{(outside vapour layer)}\,, \label{eq:staticshapes_adim} \\ 
-\frac{1}{12} \partial_{s}\left[r t^3 \partial_{s} P_v\right]&=&r\,\frac{\tilde{\mathcal{E}}}{t}\qquad\qquad \text{(vapour layer)}\,.\label{eq:general1_adim}\nonumber 
\end{eqnarray} 
When incorporating the latter (lubrication) equation in our numerical scheme, we shall aim at an incurring lubrication-approximation error as low as $O(\delta^2)$ in terms of a thin-film smallness parameter $\delta=1/t \partial_s t$. This is normally what is automatically the case for a flat substrate. However, for a curved substrate (as the pool surface here), $O(\delta)$ errors may easily occur unless care is taken. We shall eventually use the following form of that equation:   
\begin{equation}
-\frac{1}{12} \partial_{s_e}\left[\frac{r_h+r_e}{2} t^3 \left(1+\frac{1}{2}\kappa_{1e} t\right) \partial_{s_e} P_v\right]=r_h\,\frac{\tilde{\mathcal{E}}}{t\,(1+\kappa_{1e} t)}\qquad \text{(vapour layer)}\,. 
\label{eq:general1_adim_mod}
\end{equation}
Like previously with $\kappa_h$ and $\kappa_e$, the subscripts ``$h$'' and ``$e$'' are hereafter used for the geometric quantities pertaining to the drop and pool surfaces, respectively, whereas $\kappa_1$ denotes the first curvature. 
As an independent variable $s$, we have now chosen for definiteness the one along the pool surface, $s\equiv s_e$. 
The form of (\ref{eq:general1_adim_mod}) can be understood using a differential identity $\texttt{d}s_e=\texttt{d}s_h\, (1+\kappa_{1e} t)$ valid up to and including $O(\delta)$, where $\kappa_{1e} t$ is an $O(\delta)$ correction with respect to unity. As the evaporation goes on from the drop surface while $s_e$ is the arc length along the pool surface, the evaporation term on the right-hand side is accordingly modified by a factor $\texttt{d}s_h/\texttt{d}s_e$. Besides, it is $r_h$ that is used there. In contrast, for the flux in the vapour film with a symmetric lubrication profile (no-slip conditions at both surfaces), it is most precise to be based upon the middle surface (in between the drop and the pool). Hence a factor $(1+\frac{1}{2}\kappa_{1e} t)$ (with a halved correction) modifying $\partial_{s_e} P_v$ and $(r_h+r_e)/2$ for the radial coordinate on the left-hand side of (\ref{eq:general1_adim_mod}). Note that the curvature quantities $\kappa$ are here defined as positive when the surfaces are concave towards the drop (see also below). 

Equation (\ref{eq:general1_adim_mod}) together with both (\ref{eq:general2_adim}) and (\ref{eq:general3_adim}) is applied inside the vapour layer, up until the patching point located at $s_e=s_{e,\text{patch}}$. The choice of the patching point will be concretized later on. Equation (\ref{eq:staticshapes_adim}) is applied beyond the patching point to determine the equilibrium shapes of the upper part of the drop, together with (\ref{eq:general2_adim}), and of the remainder of the pool surface, together with (\ref{eq:general3_adim}).   

The following geometric relations hold: 
\begin{eqnarray}
\partial_{s_h} h=\sin\varphi_h\,,\quad 
\partial_{s_h} r_h=\cos\varphi_h\,, &\quad & 
\kappa_h= \kappa_{1h} + \frac{\sin\varphi_h}{r_h}\,,\quad
\kappa_{1h}=\partial_{s_h} \varphi_h
 \,,
\label{eq:geom_h} \\
\partial_{s_e} e=\sin\varphi_e\,,\quad 
\partial_{s_e} r_e=\cos\varphi_e\,, &\quad & 
\kappa_e= \kappa_{1e} + \frac{\sin\varphi_e}{r_e}\,,\quad 
\kappa_{1e}=\partial_{s_e} \varphi_e \,,
\label{eq:geom_e} 
\end{eqnarray}
where $\varphi$ is the angle between the tangential along the surface pointing away from the axis underneath the drop and the horizontal pointing away from the axis. For definiteness, the vapour layer thickness $t$ will be measured and the coordinate lines $s_e=\text{const}$ will be defined exactly along the orthogonals to the pool surface. Then we can also write 
\begin{equation}
h-e=t \,\cos\varphi_e \,, \quad r_e-r_h=t\,\sin\varphi_e \,.
\label{eq:geom_t}
\end{equation}

The arc length $s$ for each surface is counted from the axis underneath the drop, where we have $s_e=s_h=r_e=r_h=\varphi_h=\varphi_e=0$. We expect $\varphi_h=\pi$ at the axis at the top of the drop, whereas once again $\varphi_e=0$ at the unperturbed pool surface far away from the drop.

\subsection{Upper part of the drop}
\label{upperpart}

When solving for the upper, equilibrium part of the drop, one can get rid of the variable $s_h$ in the corresponding system of equations (\ref{eq:general2_adim}), (\ref{eq:staticshapes_adim}) and (\ref{eq:geom_h}) and render this part of the formulation in terms of an independent variable $\varphi_h$ (expected to vary monotonically in the interval of interest) and the dependent variables $h$ and $r_h$. We shall just shift the origin of $h$ to the top of the drop for later convenience. One obtains 
\begin{equation}
\partial_{\varphi_h} H = \frac{\sin\varphi_h}{\kappa_{h,\text{top}}-H-\sin\varphi_h/r_h} \,, \quad 
\partial_{\varphi_h} r_h = \frac{\cos\varphi_h}{\kappa_{h,\text{top}}-H-\sin\varphi_h/r_h} \,,
\label{eq:upper_drop} 
\end{equation}
where
\begin{equation}
h=h_\text{top}+H
\label{eq:htilde}
\end{equation}
with $h_\text{top}$ and $\kappa_{h,\text{top}}>0$ being the $h$ and curvature values, respectively, at the top of the drop. We also note that the constant $k$ appearing in (\ref{eq:general2_adim}) is hereby expressed as 
\begin{equation}
k=h_\text{top}+\kappa_{h,\text{top}} \,.
\label{eq:kkk}
\end{equation} 
The solution of the Cauchy problem (\ref{eq:upper_drop}) with the boundary condition $H=0$ and $r_h=0$ at $\varphi_h=\pi$ (removable singularity in (\ref{eq:upper_drop})) can be obtained by standard numerical methods, the integration proceeding towards $\varphi_h<\pi$. 

The parameter $\kappa_{h,\text{top}}$ is actually the one determining the size (radius $\mathcal{R}$) of the drop. The formal equation on $\kappa_{h,\text{top}}$, once a system parameter $\mathcal{R}$ is specified, is  
\begin{equation}
\max_{0<\varphi_h<\pi} r_h(\varphi_h,\kappa_{h,\text{top}}) = \mathcal{R}\,.
\label{eq:RRR}
\end{equation}
We note that in the cases when the patching point is located below the drop's equator (typically when the pool surface deformation is not too large), equation (\ref{eq:RRR}) can be solved immediately in the framework of the present equilibrium treatment of the upper part of the drop. Otherwise, equation (\ref{eq:RRR}) carries over to the vapour-layer problem to be considered in \S\ref{vaporlayer}. 

The parameter $h_\text{top}$, on the other hand, is the one determining the vertical shift of the drop, but not immediately affecting its shape. It is an unknown that carries over to the vapour-layer problem below. 

For later use (see \S\ref{vaporlayer}), to distinguish from the corresponding dependent variables in the vapour layer, we shall attribute a subscript `upper' to the solution $H(\varphi_h,\kappa_{h,\text{top}})$ and $r_h(\varphi_h,\kappa_{h,\text{top}})$ obtained in the present subsection to yield 
\begin{equation}
H_\text{upper}(\varphi_h,\kappa_{h,\text{top}}) \quad \text{and}\quad r_{h,\text{upper}}(\varphi_h,\kappa_{h,\text{top}}) \,,
\label{eq:upperpart}
\end{equation} 
presumed to be known functions in what follows and applied down to the patching point. 

\subsection{Pool surface free from the drop}
\label{freepool}

Similarly to \S\ref{upperpart}, equations (\ref{eq:general3_adim}), (\ref{eq:staticshapes_adim}) and (\ref{eq:geom_e}) describing the equilibrium shape of the pool surface beyond the patching point can be rendered in the form 
\begin{equation}
\partial_{\varphi_e} e = \frac{\sin\varphi_e}{\mathcal{P}\,\Gamma^{-1} e-\sin\varphi_e/r_e} \,, \quad 
\partial_{\varphi_e} r_e = \frac{\cos\varphi_e}{\mathcal{P}\,\Gamma^{-1} e-\sin\varphi_e/r_e} \,.
\label{eq:equilibrium_pool}
\end{equation}
Equations (\ref{eq:equilibrium_pool}) are numerically integrated towards $\varphi_e>\varphi_{e0}$ starting from numerics-adapted boundary conditions $e=C\,K_0(\mathcal{P}^{1/2}\Gamma^{-1/2} r_e)$ and $\varphi_e=-C\,\mathcal{P}^{1/2}\Gamma^{-1/2}$ $K_1(\mathcal{P}^{1/2}\Gamma^{-1/2} r_e)$ formulated 
at some numerically small $\varphi_e=\varphi_{e0}$ (numerically large $r_e=r_{e0}$). Here $K_0$ and $K_1$ are modified Bessel functions of the second kind. Such boundary conditions are inferred from (\ref{eq:staticshapes_adim}) with (\ref{eq:general3_adim}) by implying small pool surface deformations, when the curvature is given by $\kappa_e=\partial_{r_e}^2 e + r_e^{-1} \partial_{r_e} e$, and the true boundary condition $e=0$ and $\varphi_e=0$ as $r_e\to\infty$. To the hereby obtained solution $e(\varphi_e, C)$ and $r_e(\varphi_e,C)$, we append for later convenience (like in \S\ref{upperpart}) a distinctive subscript, here `free', to yield 
\begin{equation}
e_\text{free}(\varphi_e,C)  \quad\text{and}\quad r_{e,\text{free}}(\varphi_e, C) \,,
\label{eq:freepool}
\end{equation}
which are presumed to be known functions in what follows, applied down to the patching point. The constant $C$ is an unknown carrying over to the vapour-layer problem. We note that $e_\text{free}$ and $r_{e,\text{free}}$ are also functions of a system parameter $\mathcal{P}\,\Gamma^{-1}$, but an explicit argument list shall here be limited to the variables and the unknown constants. 

\subsection{Vapour layer}
\label{vaporlayer}

When solving the problem in the vapour layer, we treat $s_e$ as an independent variable in the interval $0<s_e<s_{e,\text{patch}}$. Its value at the patching point, $s_{e,\text{patch}}$, is an unknown of the problem.  
Thirteen other variables entering the formulation (\ref{eq:general2_adim}), (\ref{eq:general3_adim}), (\ref{eq:general1_adim_mod})--(\ref{eq:geom_t}) are taken as dependent (viz.~$h$, $e$, $t$, $r_h$, $r_e$, $s_h$, $\varphi_h$, $\varphi_e$, $P_v$, $\kappa_h$, $\kappa_e$, $\kappa_{1h}$, $\kappa_{1e}$). Note the possibility of formally writing $\partial_{s_h}=(\partial_{s_e} s_h)^{-1} \partial_{s_e}$ in (\ref{eq:geom_h}). For $k$ in (\ref{eq:general2_adim}), equation (\ref{eq:kkk}) is still implied. 

Overall, we have an eighth-order ODE-algebraic problem as written. The boundary conditions at $s_e=0$ are $s_h=r_e=\varphi_e=\varphi_h=\partial_{s_e} P_v=0$ ($r_h=0$ then formally following from the second algebraic equation (\ref{eq:geom_t})). At the patching point $s_e=s_{e,\text{patch}}$, we have $P_v=0$, $h=H_\text{upper}(\varphi_h,\kappa_{h,\text{top}})+h_\text{top}$, $r_h=r_{h,\text{upper}}(\varphi_h,\kappa_{h,\text{top}})$, $e=e_\text{free}(\varphi_e,C)$, $r_e=r_{e,\text{free}}(\varphi_e,C)$, where the functions marked by the subscripts `upper' and `free' are regarded known from \S\ref{upperpart} and \S\ref{freepool}, respectively. 

On the other hand, we still need a precise definition of the choice of the patching point, arbitrary within certain reasonable limits in our present scheme. We define it by setting the slope difference between the drop and pool surfaces at a sufficiently large prefixed value $\Delta\varphi$ (typically between $30^\circ$ and $90^\circ$), viz.\ $\varphi_h-\varphi_e=\Delta\varphi$ at $s_e=s_{e,\text{patch}}$, which serves as yet another boundary condition. We verify \textit{a posteriori} the results not being too sensitive to such a choice. 

Thus, the number of the boundary conditions used (eleven) can be seen to exceed by three the differential order of the problem, which is justified given that there are also five unknown constants ($s_{e,\text{patch}}$, $h_\text{top}$, $\kappa_{h,\text{top}}$, $k$, and $C$) to be determined but so far only two equations, (\ref{eq:kkk}) and (\ref{eq:RRR}), formulated for them. 

The vapour-layer problem is discretised by means of second-order finite differences at a uniform grid. The dependent variables are all defined only at the grid points themselves. The second-order ODE is discretized at the internal grid points, the first-order ODEs at the mid-points (the nonlinear terms being averaged between their values at the adjacent grid points), whereas the distributed algebraic equations hold at all grid points (due to a removable singularity in the third equations (\ref{eq:geom_h}) and (\ref{eq:geom_e}), they are replaced at the first grid point with $\kappa_h=2\kappa_{1h}$  and $\kappa_e=2\kappa_{1e}$). The boundary conditions are applied at the corresponding first or last grid points. The thereby obtained system of nonlinear algebraic equations for the values of the dependent variables at the grid points and the unknown constants, complemented yet by equations (\ref{eq:kkk}) and (\ref{eq:RRR}), is solved with the help of the \texttt{FindRoot} command in \textit{Mathematica}, which finalizes the solution for the vapour layer.  On the other hand, obtained the values of the constants ($s_{e,\text{patch}}$, $h_\text{top}$, $\kappa_{h,\text{top}}$, and $C$), the shape of the upper part of the drop is eventually given by (\ref{eq:upperpart}) applied for $\varphi_h$ in the interval $\varphi_h\big|_{s_e=s_{e,\text{patch}}}<\varphi_h\le\pi$ with the definition (\ref{eq:htilde}) taken into account, while the shape of the pool surface not covered by the drop is eventually given by (\ref{eq:freepool}) applied for $\varphi_e$ in the interval $0\le\varphi_e<\varphi_e\big|_{s_e=s_{e,\text{patch}}}$.

\section{Superhydrophobic drop computation}
\label{app:superhydrophobic}

In the present appendix, we proceed in the same non-dimensionalisation and with the same convention on hat omission as in the previous one (see the beginning of Appendix~\ref{app:numerics_full}). 

The solutions (\ref{eq:upperpart}) and (\ref{eq:freepool}) for the upper part of the drop and for the drop-free part of the pool surface obtained in the context of a Leidenfrost drop actually hold \textit{verbatim} in the context of a superhydrophobic drop. 
It is only that they must now be applied up until the contact (triple) line thereof in lieu of a patching point non-existent here. Still holding are also equations (\ref{eq:kkk}) and (\ref{eq:RRR}) and the representation (\ref{eq:htilde}). 

For the drop--pool interface (the interfacial tension $\gamma_\mathrm{d}+\gamma_\mathrm{p}$) taking place in the framework of our superhydrophobic drop, we have 
\begin{equation}
e\equiv h\,,\quad r_e\equiv r_h\,,\quad \varphi_e\equiv\varphi_h\,,\quad \kappa_e\equiv \kappa_h \,.
\label{eq:droppool_identities}
\end{equation}
The governing system of equations for this interface, a counterpart of (\ref{eq:upper_drop}) and (\ref{eq:equilibrium_pool}) for the other two interfaces, can be derived by equating the expressions (\ref{eq:general2_adim}) and (\ref{eq:general3_adim}) on account of (\ref{eq:droppool_identities}) and using geometric considerations similar to those used there. One finally arrives at 
\begin{equation}
\partial_{\varphi_h} \mathcal{H} = \frac{\sin\varphi_h}{\kappa_{h,\text{bottom}}-\frac{1-\mathcal{P}}{1+\Gamma} \mathcal{H}-\frac{\sin\varphi_h}{r_h}} \,, \quad 
\partial_{\varphi_h} r_h = \frac{\cos\varphi_h}{\kappa_{h,\text{bottom}}-\frac{1-\mathcal{P}}{1+\Gamma} \mathcal{H}-\frac{\sin\varphi_h}{r_h}} \,,
\label{eq:equilibrium_droppool} 
\end{equation}
where
\begin{equation}
h=h_\text{bottom}+\mathcal{H}
\label{eq:hhat}
\end{equation}
and
\begin{equation}
\kappa_{h,\text{bottom}}=\frac{\kappa_{h,\text{top}}+(h_\text{top}-h_\text{bottom})+ \mathcal{P}\,h_\text{bottom}}{1+\Gamma} \,,
\label{eq:kappabottom}
\end{equation}
the subscript `bottom' referring to quantities at the very bottom of the drop (at the symmetry axis). The newly introduced quantities $h_\text{bottom}$ and $\kappa_{h,\text{bottom}}$ are unknowns to be determined from the overall problem. 

The system (\ref{eq:equilibrium_droppool}) is numerically integrated starting from the boundary condition $\mathcal{H}=0$ and $r_h=0$ at $\varphi_h=0$ towards $\varphi_h>0$. To the hereby obtained solution $\mathcal{H}(\varphi_h,\kappa_{h,\text{bottom}})$ and $r_h(\varphi_h,\kappa_{h,\text{bottom}})$ we append for later distinction a subscript `lower' to yield 
\begin{equation}
\mathcal{H}_\text{lower}(\varphi_h,\kappa_{h,\text{bottom}}) \quad \text{and}\quad r_{h,\text{lower}}(\varphi_h,\kappa_{h,\text{bottom}}) \,,
\label{eq:droppool}
\end{equation} 
presumed to be known functions in what follows. 

At the contact (triple) line, we must have continuity between all the three interfaces as well as of their corresponding slopes. On account of (\ref{eq:htilde}), (\ref{eq:upperpart}), (\ref{eq:freepool}), (\ref{eq:hhat}) and (\ref{eq:droppool}), this leads to the following equations: 
\begin{equation}
	\begin{aligned}
h_\text{top}+H_\text{upper}(\varphi_\text{CL},\kappa_{h,\text{top}}) = & e_\text{free}(\varphi_\text{CL},C) = h_\text{bottom}+\mathcal{H}_\text{lower}(\varphi_\text{CL},\kappa_{h,\text{bottom}}) \,, 
\\  
r_{h,\text{upper}}(\varphi_\text{CL},\kappa_{h,\text{top}}) = & r_{e,\text{free}}(\varphi_\text{CL}, C) = r_{h,\text{lower}}(\varphi_\text{CL},\kappa_{h,\text{bottom}}) \,, 
	\end{aligned}
\label{eq:continuities_CL}
\end{equation}
where $\varphi_\text{CL}$ is the slope at the contact line (another unknown of the problem), common to all the three interfaces. 

Finally, we have a system of seven algebraic equations, (\ref{eq:kkk}), (\ref{eq:RRR}), (\ref{eq:kappabottom}), and (\ref{eq:continuities_CL}), for seven unknown constants, $k$, $h_\text{top}$, $\kappa_{h,\text{top}}$, $h_\text{bottom}$, $\kappa_{h,\text{bottom}}$, $C$, and $\varphi_\text{CL}$, which is solved numerically. Known the values of the constants and account taken of the definitions (\ref{eq:htilde}) and (\ref{eq:hhat}), the eventual shape of the superhydrophobic drop and the adjacent free pool surface is given by (\ref{eq:upperpart}) used for $\varphi_\text{CL}\le\varphi_h\le\pi$, (\ref{eq:freepool}) used for $0\le\varphi_e\le\varphi_\text{CL}$, and (\ref{eq:droppool}) used for $0\le\varphi_h\le\varphi_\text{CL}$. Known the shape, the vapor pressure (in an imaginary infinitesimal gap between the two liquids)  is given either by (\ref{eq:general2_adim}) or, what is the same, by (\ref{eq:general3_adim}), hence a known distribution $P_v(\varphi_h)$. The arc length $s_h(\varphi_h)$ along the interface between the two liquids, required in the formulation of \S\ref{sec:extend}, can be determined by \textit{a posteriori} integrating $\partial_{\varphi_h} s_h=\sqrt{(\partial_{\varphi_h} r_{h,\text{lower}})^2+(\partial_{\varphi_h} \mathcal{H_\text{lower}})^2}$ and implying $s_h=0$ at the symmetry axis. In this way, we can express $P_v$ and $r_{h,\text{lower}}$ as functions of $s_h$, which we symbolically rewrite below in the notation to be used in \S\ref{sec:extend}: 
$$\hat{P}_v(\hat{s})\,,\quad \hat{r}(\hat{s})\,,$$
i.e. with a hat restored (cf.~the beginning of Appendix~\ref{app:numerics_full}) and subscripts `$h$' and `lower' dropped. The value of this arc length at the contact line, $s_h(\varphi_\text{CL})$, will likewise be denoted as $\hat{s}_\text{CL}$ in \S\ref{sec:extend}.

\end{document}